\documentclass{elsart5p}

\usepackage{natbib}

\usepackage{graphicx}

\usepackage{amssymb}

\usepackage{color}
\usepackage{rotating}

\usepackage{multirow}
\usepackage{array}
\usepackage{natbib}

\begin{document}

\begin{frontmatter}



\title{Supernova progenitors and iron density
  evolution from SN rate evolution measurements}  


\author[paris]{G. Blanc\corauthref{cor1}},
\ead{blanc@apc.univ-paris7.fr}
\author[padova]{L. Greggio}
\ead{laura.greggio@oapd.inaf.it}
\address[paris]{Laboratoire APC,
10, rue Alice Domon et L\'eonie Duquet,
75205 Paris Cedex 13, France.}
\address[padova]{INAF Osservatorio Astronomico di Padova, 
Vicolo dell'Osservatorio
 5, 35122 Padova, Italy.}
\corauth[cor1]{Tel: +33 1 57 27 60 51 / Fax: +33 1 57 27 60 71}

\begin{abstract}

Using an extensive compilation of literature supernova rate data we
study to which extent its evolution constrains the star formation
history, the distribution of the type Ia supernova (SNIa) progenitor's
lifetime, the mass range of core-collapse supernova (CCSN)
progenitors, and the evolution of the iron density in the field.

We find that the diagnostic power of the cosmic SNIa rate on their
progenitor model is relatively weak. More promising is the use of the
evolution of the SNIa rate in galaxy clusters.  We find that the CCSN
rate is compatible with a Salpeter IMF, with a minimum mass for their
progenitors $\gtrsim 10\ \mathrm{M}_{\odot}$.  We estimate the
evolution in the field of the iron density released by SNe and find
that in the local universe the iron abundance should be $\sim 0.1$
solar. We discuss the difference between this value and the iron
abundance in clusters.

\end{abstract}


\end{frontmatter}




\section{Introduction}

As the last stage of the evolution of stars in a wide range of masses,
supernovae play a unique role in the cycling of matter in galaxy
evolution.  They release in the interstellar medium all of the iron
available in the Universe, which is produced (mainly as radioactive
Nickel 56) in the explosion itself \citep{woosley2002}. The supernova
rate (SNR) is then closely related to the rate of Fe enrichment of
galaxies and of the universe in general.

There are two main families of supernovae, the ``Core-Collapse
Supernovae'' (CCSN) which regroup into the observational types II, Ib
and Ic, and the type Ia SNe (SNIa), which are the product of the
thermonuclear explosion of a white dwarf (WD), accreting matter from a
companion.

CCSN are the result of the death of massive stars. Modeling the
explosion is rather difficult due to the complicated physical
processes taking place in the extreme conditions of the collapsing Fe
core \citep[see e.g.][]{woosley1986,janka2007}. This reflects into an
intrinsic difficulty to predict the exact mass range of the
progenitors of the different CCSN kinds, as well as of the amount of
radioactive nickel released to the interstellar medium by each event.
However, there is no doubt that the lifetimes of CCSN progenitors are
short, less then a few 10 Myr, which implies that the CCSN rate
closely follows the star formation rate. 

The main frame of the SNIa theoretical paradigm needs a CO white dwarf
in a close binary system. The primary star results from the evolution
of an intermediate mass star ($M~\lesssim~8\ \mathrm{M}_{\odot}$)
which develops a degenerate CO core. When, due to evolution, the
secondary expands and fills its Roche Lobe, the CO WD may accrete and
grow in mass, and, if the Chandrasekhar limit is reached, C is ignited
under degenerate conditions, initiating a thermonuclear
explosion. This is the Single Degenerate (SD) channel for SNIa's
production \citep[see e.g.][]{hachisu1999}. Conversely, if efficient
growth of the CO WD does not occur, the envelope of the secondary is
dispersed in the interstellar medium, leaving behind two close
WDs. The binary system has a second chance to give rise to a SNIa
event, since, due to the emission of gravitational waves radiation,
orbital energy is lost and the two WDs eventually merge.  If the total
mass of this Double Degenerate (DD) system exceeds the Chandrasekhar
limit a SNIa may be produced \citep{iben1984}. While the modeling of
the binary system evolution from its birth to the final explosion is
subject to many relatively uncertain prescriptions, the
nucleosynthetic product of the explosion seems more robustly assessed,
and it consists of mainly Fe peak elements, and intermediate (Si
group) mass elements \citep{iwamoto1999}.  The time delay between the
birth of the binary system and the final explosion spans a wide range,
from a few $10^7$~yrs, corresponding to the nuclear evolutionary
lifetime of the more massive systems, up to a Hubble time or more,
especially for the DD systems born with the larger separations.  As a
consequence, the scaling of the SNIa rate with time will reflect not
only the star formation history of the parent stellar system, but also
the distribution of the delay times.  

In this paper we analyze the constraints on the progenitors of CCSN,
on the distribution of the delay times of SNIa, and on the cosmic
history of star formation which can be derived from the measurements
of the cosmic SN rates of the two kinds.  Previous studies on this
subject include
\citet{jorgensen1997,sadat1998,dahlen1999,dahlen2004,forster2006}.
With respect to these, we try to test the conclusions versus the
choice of relevant parameters, by considering a variety of options for
the distribution of the delay times and for the SFH.

The paper is organized as follows: in Sec.~\ref{sec:measurements}, SNR
measurements are reviewed; in Sec.~\ref{sec:modelisation}, the various
theoretical ingredients hidden in the SNR evolution are dissected; in
Sec.~\ref{sec:results} we discuss the fit of the models to the
observed cosmic SNIa rate in the field and in clusters, and to the
observed cosmic CCSN rate. In Sec.~\ref{sec:iron} we discuss the
implications of our results on the evolution of the Fe density in the
universe.  Conclusions are drawn in Sec.~\ref{sec:conclusions}.

We use the concordance cosmological model $(\Omega_{M_{\circ}},
\Omega_{\Lambda_{\circ}}) = (0.3, 0.7)$, with a Hubble constant set to
H$_{\circ} = 70h_{70}~\mathrm{km}\ \mathrm{s}^{-1}
\mathrm{Mpc}^{-3}$. We use the SNu or ``\textit{SuperNova Unit}'', 1
SNu = 1~SN/$10^{10} \mathrm{L}_{B \odot}/\mathrm{century}$.

\section{Supernova rate measurements}
\label{sec:measurements}

Together with the first high-redshift supernova searches, the 1990's
have seen the first type Ia rate measurement beyond the local universe
\citep{pain1996} at $z \sim 0.4$, using for the first time a
homogeneous sample of CCD discovered SNe.  Other supernova searches or
surveys have provided an estimate of the rate at various redshifts:
EROS -- \citet{hardin2000,blanc2004}, SCP -- \citet{pain2002}, SDSS --
\citet{madgwick2003}, HZT -- \citet{tonry2003}, \citet{barris2006},
GOODS -- \citet{dahlen2004,kuznetsova2008}, SNLS -- \citet{neill2006},
Subaru Deep Field -- \citet{poznanski2007}, STRESS --
\citet{botticella2008}. Tab.~{\ref{tab:otheresultsIa}} summarizes all
the present SNIa rate measurements including the rate in the local
universe derived by combining the \citet{cappellaro1999}
determinations for all galaxy kinds, for homogeneity with the high
redshift measurements.

Several attempts have been made to measure the SNIa rate in
galaxy clusters. This approach is interesting since the galaxy
population in cluster seems rather homogeneous and old. First
measurement in high redshift clusters has been made by
\citet{gal-yam2002}, with little statistics. \citet{reiss2000} used
the \textit{Mount Stromlo Abell Cluster SN Search} to derive a rate
within clusters. The latest measurement of this kind has been
published by \citet{sharon2007}, where a summary
of all current SNIa rate measurements in clusters is presented.

First high-redshift measurements of the CCSN rate have been published
by two teams almost simultaneously, the STRESS survey
\citep{cappellaro2005} and the GOODS survey \citep{dahlen2004}. The
result from STRESS survey by \citet{botticella2008} supersedes the one
of \citet{cappellaro2005}.  Tab.~{\ref{tab:otheresultsCC}} summarizes
these measurements. The high-redshift CCSN sample used to derive the
rate is not completely based on spectroscopic confirmation, both for
\citet{cappellaro2005,botticella2008} and \citet{dahlen2004}
measurements. Another source of uncertainty may come from the
extinction within the host, since CCSN are known to occur shortly
after their birth in highly dusty star formation regions of
galaxies. Measurements by \citet{cappellaro1999} are not corrected for
extinction. \citet{dahlen2004} attempted to make such a correction by
simulating the host dust distribution according to the disk
inclination, following the prescription by \citet{hatano1998}. Their
corrected rates (as used in this paper) are a factor of $\sim$ 2
higher than the uncorrected. We notice, however, that such a
correction is still uncertain as discussed by the authors.
\citet{botticella2008} did a modeling of host galaxy extinction
following the recipe of \cite{riello2005}; their rate measurements
take it into account both for SNIa and CCSN.

\section{Modelization of the supernova rate evolution}
\label{sec:modelisation}

The SN rate ($\mathcal{R}_{\mathrm{SN}}$) as a function of time can be
modeled as the convolution of the SFH ($\Psi$) with the
distribution function ($f_{\mathrm{SN}}$) of the delay times ($\tau$):
\begin{equation}
\mathcal{R}_{\mathrm{SN}}(t) = 
\int_{\tau_{\mathrm{min}}}^{\mathrm{min}(t,\tau_{\mathrm{max}})}k_{\Gamma}(t-\tau) \cdot 
\Psi(t-\tau) f_{\mathrm{SN}}(\tau) A_{\mathrm{SN}}(t-\tau) d\tau
\label{eqn:ratevstime}
\end{equation}
where $t$ is the time elapsed since the beginning of star formation in
the system under analysis; $k_{\Gamma}$ is the number of stars per
unit mass of the stellar generation with age $\tau$;
$\tau_{\mathrm{min}}$ and $\tau_{\mathrm{max}}$ bracket the range of
possible delay times; $A_{\mathrm{SN}}$ is the number fraction of
stars from the stellar generation born at epoch $(t-\tau)$ that end up
as SN. 
For core-collapse events, the progenitor life-time is very short,
($\tau < 0.05\ \mathrm{Gyr}$), so that
Eq.~(\ref{eqn:ratevstime}) can be approximated as:
\begin{equation}
\mathcal{R}_{\mathrm{CC}}(t) = k_{\Gamma}A_{\mathrm{CC}} \cdot \Psi(t)
\label{eqn:ccvstime}
\end{equation}
with the SFR and number of SN per unit mass of the parent stellar
generation ($k_{\Gamma} A_{\mathrm{CC}}$) evaluated at the
current epoch.  For type Ia events a very wide range of delay times is
predicted from stellar evolution, with $\tau_{\mathrm{min}}$ of a few
tens of Myr, and $\tau_{\mathrm{max}}$ of the order or one Hubble
time, or more.  We can still introduce a little simplification to
Eq.~(\ref{eqn:ratevstime}) by assuming that the number of SNIa from 1
M$_{\odot}$ stellar generation does not depend on time and write:
\begin{equation}
\mathcal{R}_{\mathrm{Ia}}(t) = k_{\Gamma} A_{\mathrm{Ia}} \cdot
\int_{\tau_{\mathrm{min}}}^{\tau_{\mathrm{max}}} \Psi(t-\tau)
f_{\mathrm{Ia}}(\tau) d\tau
\label{eqn:rateIavstime}
\end{equation}
which shows that the SNIa rate evolution depends both on the SFH and
on the distribution of delay times.  

Both $A_{\mathrm{CC}}$ and $A_{\mathrm{Ia}}$ can be evaluated from
stellar evolution, including a choice for the IMF. However, these
theoretical values depend on assumption on the mass range of the
progenitors. In addition, $A_{\mathrm Ia}$ depends on the distribution
of binary separations and mass ratios, the outcome of the mass
exchange phases, and of the final accretion on top of the
WD. Therefore, we prefer to determine these quantities empirically
from the fit of the SN rates, and compare the derived values to the
theoretical expectations.

\subsection{The Star Formation History}
\label{sec:sfh}

\begin{figure*}[!htb]
\centering
\includegraphics[width=17cm]{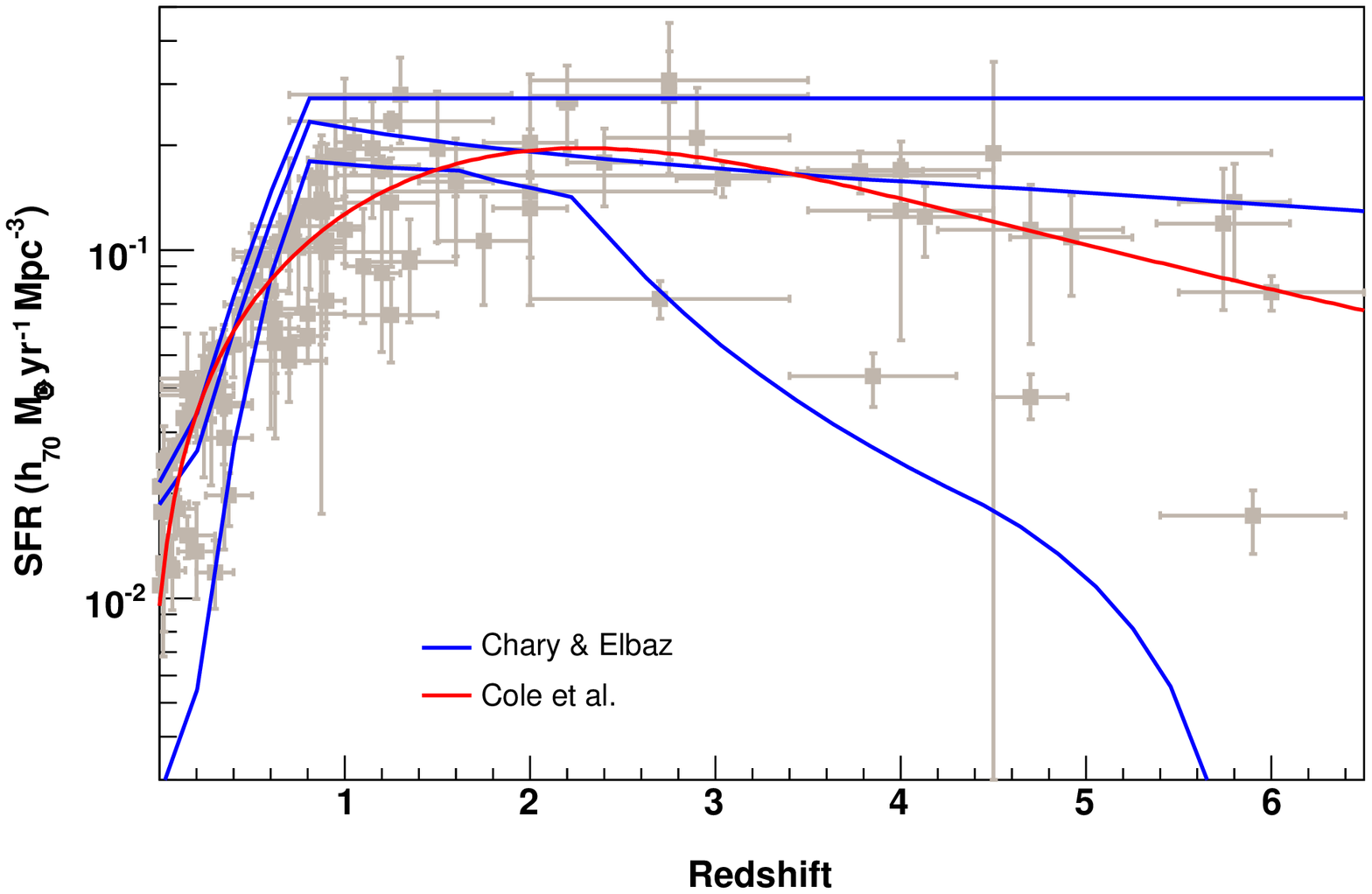}
\caption{Measured cosmic SFR as compiled by \citet{hopkins2004} and
  \citet{hopkins2006}. Models are from \citet{chary2001}, extrapolated
  for $z > 4.5$ and \citet{cole2001}.}
\label{fig:sfrz}
\end{figure*}

The SFH, $\Psi(t)$, is currently measured up to $z \sim 6$ (with some
estimates up to $z \sim 10$, \citet{bouwens2005}) from galaxy surveys,
using luminosity tracers and exploiting correlations of SFH with UV,
H$_{\alpha}$, FIR luminosities (see e.g. \citet{kennicutt1998b}). It
has been pointed out that short-wavelength measurements lead to an
underestimate of the SFR because of dust absorption at
high-redshift. Then FIR luminosity which is less affected by the dust
obscuration might be a better indicator \citep{chary2001}.
Fig.~\ref{fig:sfrz} shows the compilation of SFR measurement by 
\citet{hopkins2004} and \citet{hopkins2006}, who rescale the literature
data to derive a homogeneous sample, with the same cosmology, IMF and
correction for absorption.
Although the data points generally describe a consistent picture,
there are large discrepancies at given redshift, which likely follow
from the different systematics in the heterogeneous sample (see
discussion in \citet{hopkins2006}). As a result, several laws for
$\Psi(t)$ are consistent with the data.  For our exercise, we consider
the following two options: \citet{chary2001} SFH -- hereafter CE, a
model derived from far-IR SFR measurements under the form of
acceptance interval, which we implement with its average values at
each redshift (middle blue curve in Fig.~\ref{fig:sfrz}); and the
function introduced by \citet{cole2001} and used by
\citet{hopkins2006} to fit their SFR measurements compilation:
$\rho_{\star}(z) = 0.7h_{70} (a+bz)/(1+(z/c)^d)$.  Our fit to the data
based on a Salpeter IMF gives: $a = 0.0134, b=0.175,
c=2.93,d=3.01$. The two SFH mainly differ in their peak in redshift,
which is $\simeq$ 0.8 and 2 respectively for CE and Cole. They also
correspond to two different values for the total gas mass cycled into
stars, with CE being a factor 1.3 larger than Cole's.

Since $\Psi(t)$ is derived from sampling the massive stars, a large
extrapolation factor is implied in the derivation of the total SFR,
which depends on the assumed IMF. In addition, the conversion factor
between UV luminosity and star formation rate in massive stars depends
on the slope of the IMF in the high mass range. As long as this slope
is similar to the Salpeter's the conversion factor does not change,
and the $\Psi(t)$ derived for Salpeter IMF can be rescaled to another
IMF with a different proportion of low mass stars. In our computations
we adopt a Salpeter IMF.

\subsection{The Distribution of the delay times}

The distribution function of the delay times describes
the proportion of early and late explosions past an instantaneous burst of star
formation. Indeed, for such a case Eq.~(\ref{eqn:rateIavstime}) becomes:
\begin{equation}
\mathcal{R}_{\mathrm{Ia}}(t) = k_{\Gamma} A_{\mathrm{Ia}} \cdot
\mathcal{M}_{\mathrm{B}} \cdot
f_{\mathrm{Ia}}(\tau = t) 
\label{eqn:rateIaburst}
\end{equation}
where $\mathcal{M}_{\mathrm{B}}$ is the total gas mass turned into
stars in the burst. The theoretical $f_{\mathrm{Ia}}$ function is
uncertain because of various reasons: on the one hand there exist
different classes of potential SNIa progenitors, on the other, the
delay time distributions of each class depend on the distribution of
binary parameters, and on the effect on them of the mass exchange
phases which occur during the evolution of close binary systems
\citep[see][ for a discussion]{greggio2005}. In the literature
different formulations can be found for $f_{\mathrm{Ia}}$, some
stemming from stellar evolution prescriptions
\citep[e.g.][]{sadat1998,yungelson2000,han2003,han2004,greggio2005},
others from convenient analytical parametrizations
\citep[e.g.][]{madau1998,mannucci2005,scannapieco2005}, others from
the fit of the observed time behaviour of the SNIa rate
\citep[e.g.][]{strolger2004}. For this paper we have selected a few of
these proposed distribution function of the delay times to evaluate
their impact on the interpretation of the cosmic evolution of the SNIa
rate, namely the \citet{madau1998} formulation, a selection of
\citet{greggio2005} models, and a selection of the
\citet{han2003,han2004} simulations. The first are parametric analytic
formulae which allow the exploration of a wide range of shapes; the
third is meant to illustrate the result of specific predictions from
binary population synthesis.

\begin{figure*}[!htb]
\centering
\includegraphics[width=17cm]{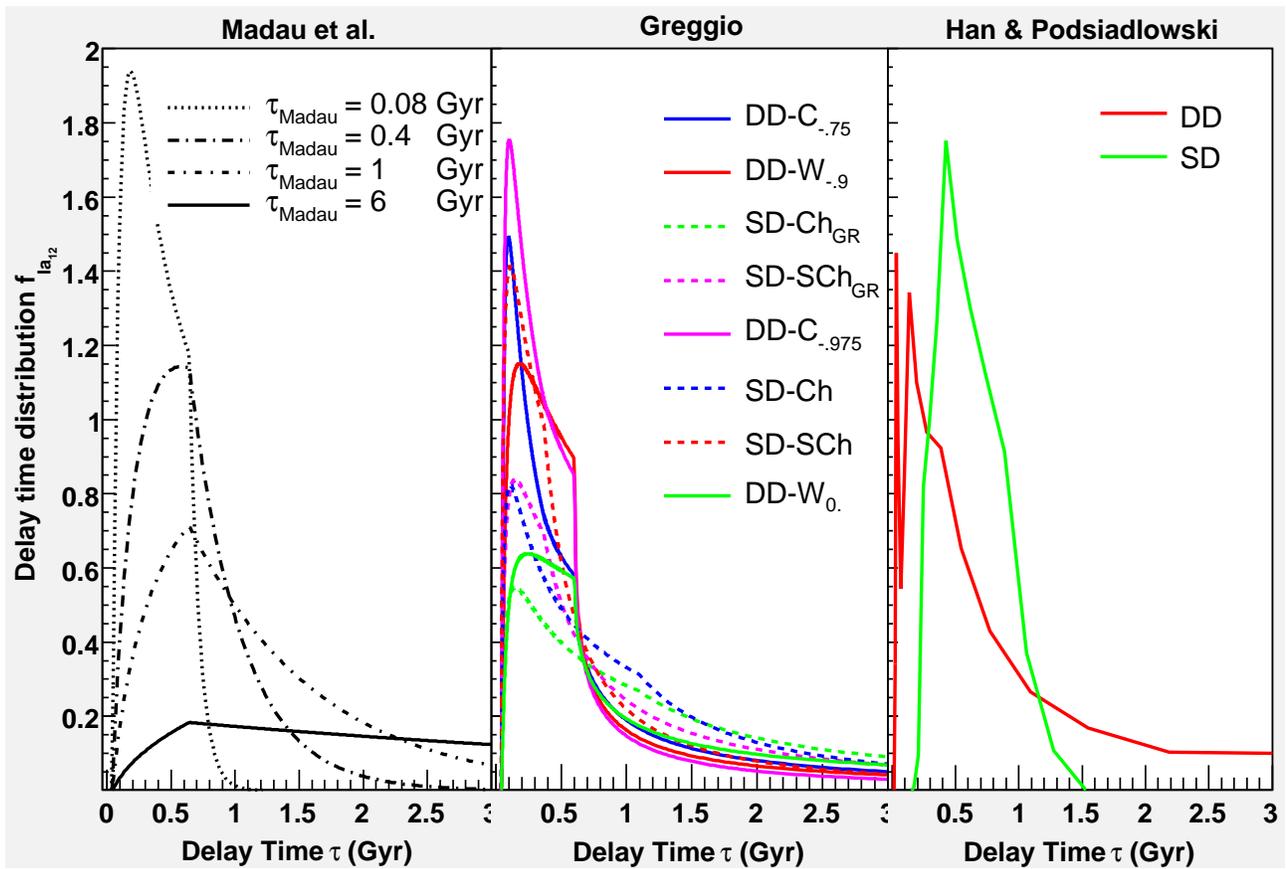}
\caption{Delay time distributions used: Madau et al. (left), Greggio
  (middle) and Han and Podsiadlowski (right). They are all normalized
  over the range 0-12 Gyr.}
\label{fig:fIas}
\end{figure*}

Left panel in Fig. ~\ref{fig:fIas} shows examples of the
\citet{madau1998} distributions, i.e.
\begin{equation}
f_{\mathrm{Ia}}(\tau) \propto
\int_{\mathrm{max}({m_{\mathrm{Ia}_{\mathrm{min}}}},(\tau/10)^{-0.4})}^{m_{\mathrm{Ia}_{\mathrm{max}}}} 
\exp \left(- \frac{\tau-t_{\mathrm{ms}}}{\tau_{\mathrm{Madau}}}
\right) m^{-2.35} dm
\label{eqn:fIaMadau}
\end{equation}
where $t_{\mathrm{ms}} = 10/m^{2.5}$ is the main sequence lifetime (in
Gyr) of a star of mass $m$ (in $\mathrm{M}_{\odot}$), and
$\tau_{\mathrm{Madau}}$ is a characteristic delay time, considered as
a free parameter.  $m_{\mathrm{Ia}_{\mathrm{min}}}$ and
$m_{\mathrm{Ia}_{\mathrm{max}}}$ bracket the mass range of SNIa
progenitors, for which we have adopted 3 and 9 $\mathrm{M}_{\odot}$
respectively. As $\tau_{\mathrm{Madau}}$ increases, the distribution
of the delay times becomes wider, still with most of the explosions
occurring at early delays. For a long $\tau_{\mathrm{Madau}}$ the
distribution becomes flat.

The central panel in Fig.~\ref{fig:fIas} shows the selected models
from \citet{greggio2005}. These are analytic delay time distribution
functions based on general arguments on the evolutionary behaviour of
stars in binary systems. The different SD models, shown as dotted
lines, correspond to different schemes for the derivation of the
distribution of primary and secondary mass in the SNIa progenitors,
and to different options for the explosion (Chandrasekhar or
Sub-Chandrasekhar). All cases adopt a flat distribution of the mass
ratios (i.e. $\mu=0$). The DD models, shown as solid lines, are
characterized by different assumptions about the result of the first
common envelope evolution (WIDE or CLOSE systems), and different
slopes for the distribution of the separations of the DD system at
birth.  This is described by the $\beta$ parameter, whose definition
is different in the CLOSE ($\beta_{\mathrm g}$) and WIDE DD
($\beta_{\mathrm a}$) schemes.  Larger values of these parameters
imply flatter slopes of the distribution of the delay times at late
epochs. All DD models in Fig.~\ref{fig:fIas} assume that the minimum
mass of the secondary component in SNIa progenitor system is $\simeq$
2.5 M$_{\odot}$, which corresponds to a maximum nuclear delay of
$\tau_{\mathrm n,x}=0.6$ Gyr. We refer the reader to the
\citet{greggio2005} paper for a more detailed description of the
parameters.  All of the distribution functions in Fig.~\ref{fig:fIas}
exhibit an early rise, followed by a populated peak of variable width,
followed by a tail of late explosions.  The proportion of
early-to-late epoch explosions is highly variable among the different
models.

In Fig.~\ref{fig:fIas} we also plot our adopted models from
\citet{han2003,han2004} in the right panel. The Han DD model (red),
refers to double degenerate progenitors, with a standard
parametrization for the common envelope evolution
($\alpha_{\mathrm{CE}}=\alpha_{\mathrm{th}}=1,
\alpha_{\mathrm{RLOF}}=0.5$ in \citet{han2003} notation).  The Han SD
model (green) is a single degenerate, Chandrasekhar model with initial
secondary masses between $\sim 2$ and 3.5 M$_\odot$. These limits
follow from the requirement of stability during the mass transfer from
the donor to the WD, so that the latter component can reach the
Chandrasekhar mass. The direct consequence of the lower limit on the
secondary mass is that this SD model lacks the late delay times tail
in the $f_\mathrm{Ia}$ function.

\section{Results}
\label{sec:results}

In this section we explore which constraints can be derived by the
analysis of the cosmic SN rate, by applying Eqs.~(\ref{eqn:ccvstime})
and ~(\ref{eqn:rateIavstime}). Fits of SN rate are done by a $\chi^2$
method, providing an estimate of goodness-of-fit as a ``$\chi^2$
probability''. We note that the supernova rate statistics is
poissonian, which converges toward gaussian statistics for big
numbers. Although most of SN rate used in this paper are based on
small numbers of events, we assume that we can use the $\chi^2$
fitting method.  Systematic errors (when available), are taken into
account by adding them quadratically to the statistical error.  For
each fit we derive a value for the $k_{\Gamma} A_{\mathrm{SN}}$
factor, \textit{i.e.} the number of SN per unit mass\footnote{This
is the mass of the stellar population at zero age, which is larger
than its mass at any subsequent age because of the mass return from
stars.}  from the parent stellar population needed to account for the
observed level of the rates.

The SNIa rates by \citet{dahlen2004} have been recently re-computed by
\citet{kuznetsova2008} with a more sophisticated method. For our fit
we adopt the most up-to-date values. 

Similarly, for the STRESS survey, we adopt the CCSN rate in
\citet{botticella2008} which updates the \citet{cappellaro2005} value
on the basis of a larger SN sample.

\subsection{SNIa rate and SNIa progenitors}
\label{sec:ressn1a}

\begin{figure*}[!htb]
\centering
\includegraphics[width=17cm]{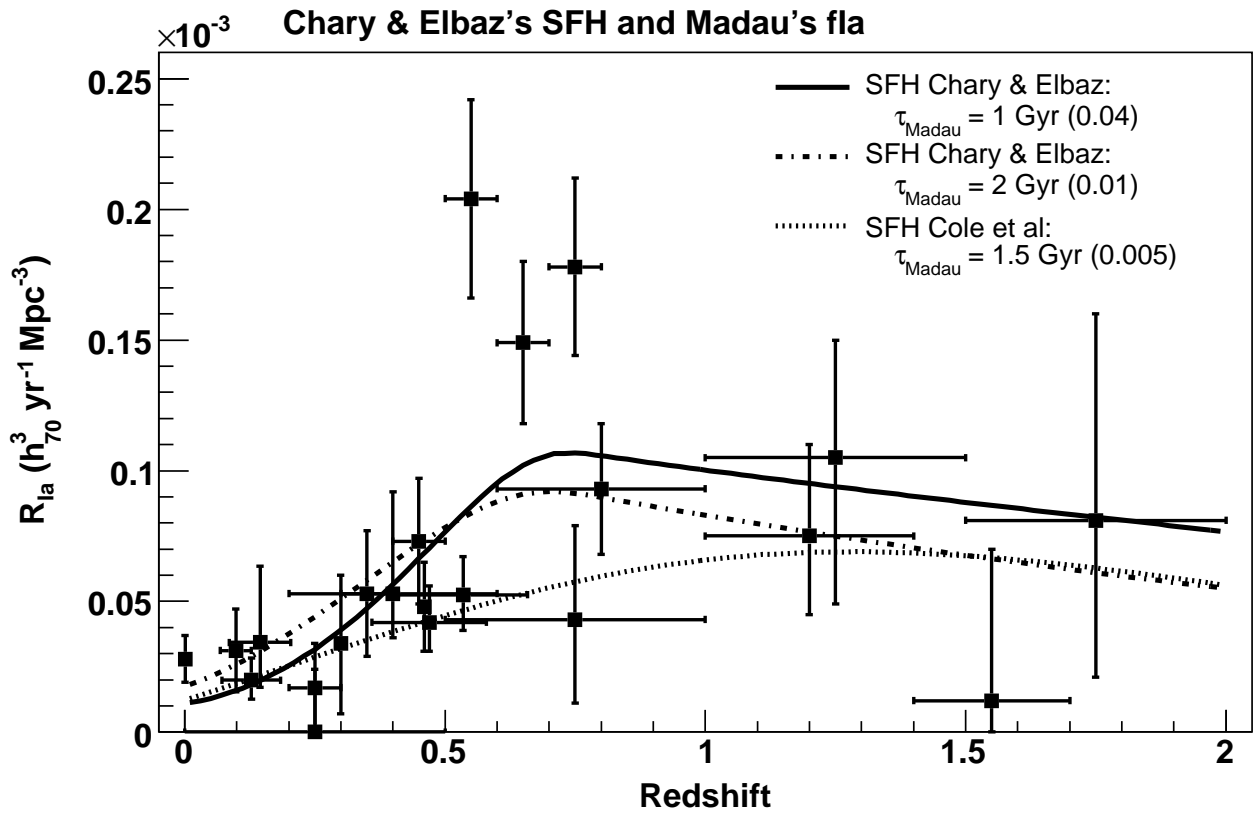}
\caption{SNIa rate measurements fitted with Madau's delay time
distribution and both CE and Cole SFH. The value of the fitted Madau's
$\tau$ parameter (here $\tau_{\mathrm{Madau}}$) is indicated. $\chi^2$
fit probabilities are shown in parenthesis: best fits are
$\tau_{\mathrm{Madau}} = 1\ \mathrm{Gyr}$ for CE SFH, and
$\tau_{\mathrm{Madau}} = 1.5\ \mathrm{Gyr}$ for Cole SFH.}
\label{fig:SNRIa_z_Madau}
\end{figure*}

\begin{figure*}[!htb]
\centering
\includegraphics[width=17cm]{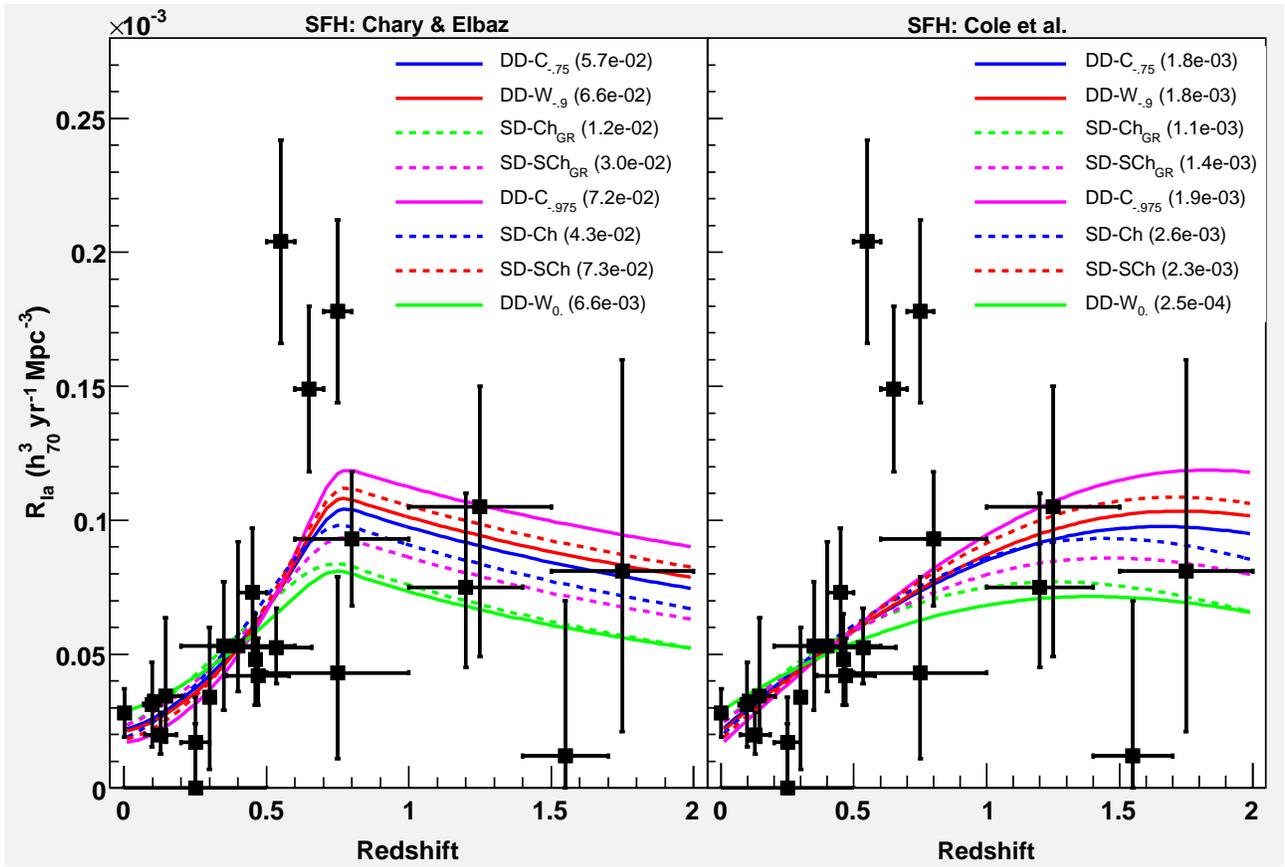}
\caption{SNIa rate measurements fitted with Greggio's delay time
  distribution and SFH from CE (left) and Cole et
  al. (right). $\chi^2$ fit probabilities are shown in parenthesis.}
\label{fig:SNRIa_z_Greggio}
\end{figure*}

\begin{figure*}[!htb]
\centering
\includegraphics[width=17cm]{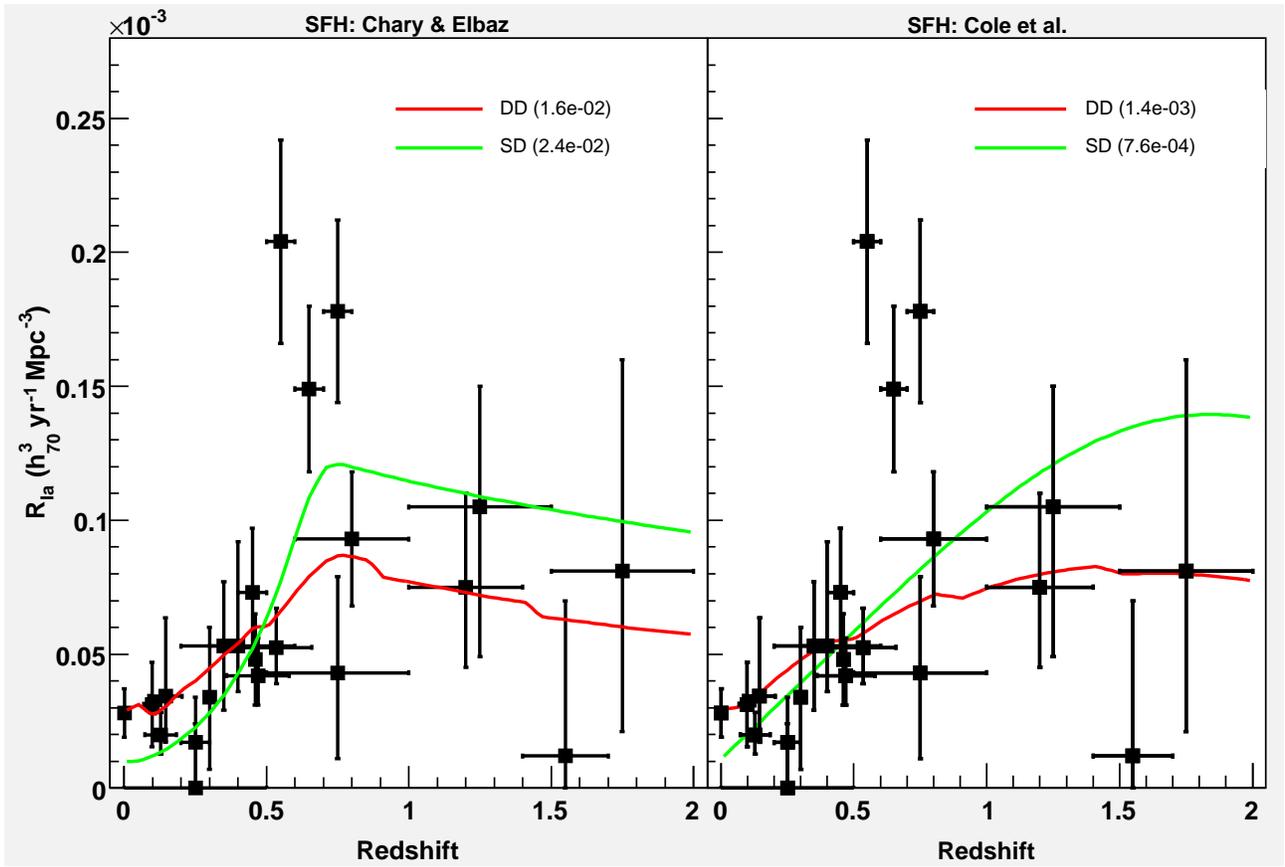}
\caption{SNIa rate measurements fitted with Han and Podsiadlowski's
  delay time distribution and SFH from CE (left) and Cole et
  al. (right). $\chi^2$ fit probabilities are shown in parenthesis.}
\label{fig:SNRIa_z_Han}
\end{figure*}

Fig.~\ref{fig:SNRIa_z_Madau} illustrates our fit with the Madau's
functions. The best case is obtained with $\tau_{\mathrm{Madau}} = 1\
\mathrm{Gyr}$, corresponding to a very peaked distribution of the
delay times (cf.~Fig.~\ref{fig:fIas}), in combination with CE
SFH. \citet{cole2001}'s cosmic SFR is excluded by about 3-sigma
($\mathcal{P}(\chi^2) = 5\cdot 10^{-3}$). Similarly,
Fig.~\ref{fig:SNRIa_z_Greggio} shows the fit to the data with the
\citet{greggio2005} distribution functions. The \citet{cole2001}'s SFH
do not provide a very good fit, while they all get a significant fit
probability (over a few percents) when CE's SFH is
adopted. Fig.~\ref{fig:SNRIa_z_Han} shows the fit based on our Han and
Podsiadlowski selected models. Again the best fit is obtained in
combination with CE SFH.  The main difference between the two SFH is
the position of maximum, $z_{\mathrm{max}} \sim 0.8$ for
\citet{chary2001}, and $z_{\mathrm{max}} \sim 2$ for \citet{cole2001}.
We conclude that the SNIa rate data favor a maximum of SFH at late
epoch in the history of the universe.

\citet{forster2006} obtained a similar conclusion with different
ingredients: the typical delay time of SNIa progenitors inferred from
the cosmic evolution of their rate depends on the SFH adopted.
Inspection of Fig.~\ref{fig:SNRIa_z_Greggio}, however, shows that
even for a given cosmic SFH, the model for the SNIa progenitors is not
well constrained, with all models running very close to each other.
Although an adopted cosmic SFH allows to solve for one most probable
delay time, this does not tell much about the progenitors of SNIa,
whose delay times are spread over a wide distribution.

In general, the diagnostic power of the cosmic SNIa rate is relatively
weak, due to the fact that these data are averages of the SNIa rate in
all galaxy types, over the SFH of the whole universe.  In other words,
it results from the convolution of the $f_{\mathrm Ia}$ function with
a wide $\Psi(t)$ function. A better way to constrain the SNIa model
would consist in considering the SNIa rate evolution as function of
galaxy type or in galaxy clusters, which are characterized by a
relatively narrow $\Psi(t)$ \citep{gal-yam2002}.  This is attempted in
the next section.

The values obtained for the $k_{\Gamma} A_{\mathrm{Ia}}$ factors
(where $k_{\Gamma} = 2.83\ \mathrm{M}_{\odot}^{-1}$ for a Salpeter IMF)
from fitting the cosmic SNIa rate with Greggio's distributions range
in $[0.45-0.55]\cdot 10^{-3}\ \mathrm{M}_{\odot}^{-1}$ and
$[0.58-0.71]\cdot 10^{-3}\ \mathrm{M}_{\odot}^{-1}$ when adopting CE
and Cole SFH respectively. The difference reflects the different total
star formation for the two options, with CE providing more stellar
mass than Cole's, so that the level of the SNIa rate is recovered in
combination with a lower realization probability of the event.  The
fit obtained with Han distributions and CE SFH yields similar values:
$k_{\Gamma} A_{\mathrm{Ia}} = [0.33;0.53]\cdot 10^{-3}\
\mathrm{M}_{\odot}^{-1}$ for DD and SD models respectively. Notice
that the fits obtained with Han's distributions in
Fig.~\ref{fig:SNRIa_z_Han} bracket the curves obtained with Greggio's
distributions.  We compare these derived factors to the number of
stars per unit mass of the parent stellar population with mass between
3 and 9 M$_\odot$, which is an indicative range of the mass of SNIa
progenitors (Umeda et al. 1999).  Since Salpeter IMF provides $0.022\
\mathrm{M}_{\odot}^{-1}$ such stars, the $k_{\Gamma} A_{\mathrm{Ia}}$
factors require that about 2-3 \% of them give rise to a SNIa event.

We notice the large discrepancy between rates measured by
\citet{barris2006} at $z\gtrsim 0.5$ and by others: at $z\sim 0.55$
the discrepancy with the \citet{pain2002} measurement reaches more
than 3$\sigma$. Since the \citet{barris2006} measurements are not
fully based on spectroscopical confirmation and do not include
systematics errors, we also perform the fit excluding the 3 discrepant
points. The resulting $k_{\Gamma} A_{\mathrm{Ia}}$ factors change for
less than 20 \% but the quality of fit drastically increases, 91 to 97
\% for Greggio models combined with CE SFH, and 54 to 70 \% for
Greggio with Cole SFH. This favors Han's DD model (97 \% and 74 \% for
CE and Cole SFH respectively), while their SD model gets a 50 \% fit
probability.  The quality of the fits has become so close that we
cannot conclude for one or the other SFH model. In this respect, we
notice that rate by \citet{dahlen2004} at $z\sim 0.8$ is similar to
the \citet{barris2006} value, but its redetermination by
\citet{kuznetsova2008}, and the new data by \citet{poznanski2007}
favor lower SNIa rates at these redshifts.

\subsection{SNIa rate in galaxy clusters}

The galaxy population in clusters is dominated by Early Type Galaxies
(ETGs), which, according to current understanding, contain mostly old
stars \citep[see][ for a review]{renzini2006}.  This is so not only in
the local universe, but at least all the way to $z \sim 1$
\citep{tanaka2005}. Therefore, it is generally expected that the
evolution of the SNIa rate in clusters of galaxies basically traces
the evolution of the SNIa rate in ETGs, which closely scales as the
distribution function of the delay times, as long as their SFH can be
described as a short initial burst \citep{gal-yam2004}.

In a cluster, the SNIa rate in SNu,
$\mathcal{R}_{\mathrm{Ia},\mathrm{SNu}}$, is given by the sum over the
cluster members of their individual rates in SNu,
$R_{\mathrm{Ia},\mathrm{SNu}}$, weighted by their contribution to the
total B luminosity, $f_{B}$:

\begin{equation}
\mathcal{R}_{\mathrm{Ia},\mathrm{SNu}} = \sum_{i}
R_{\mathrm{Ia},\mathrm{SNu},i} \cdot f_{B,i}.
\label{eqn:rIaclsnu}
\end{equation}

\begin{figure*}[!htb]
\centering
\includegraphics[width=17cm]{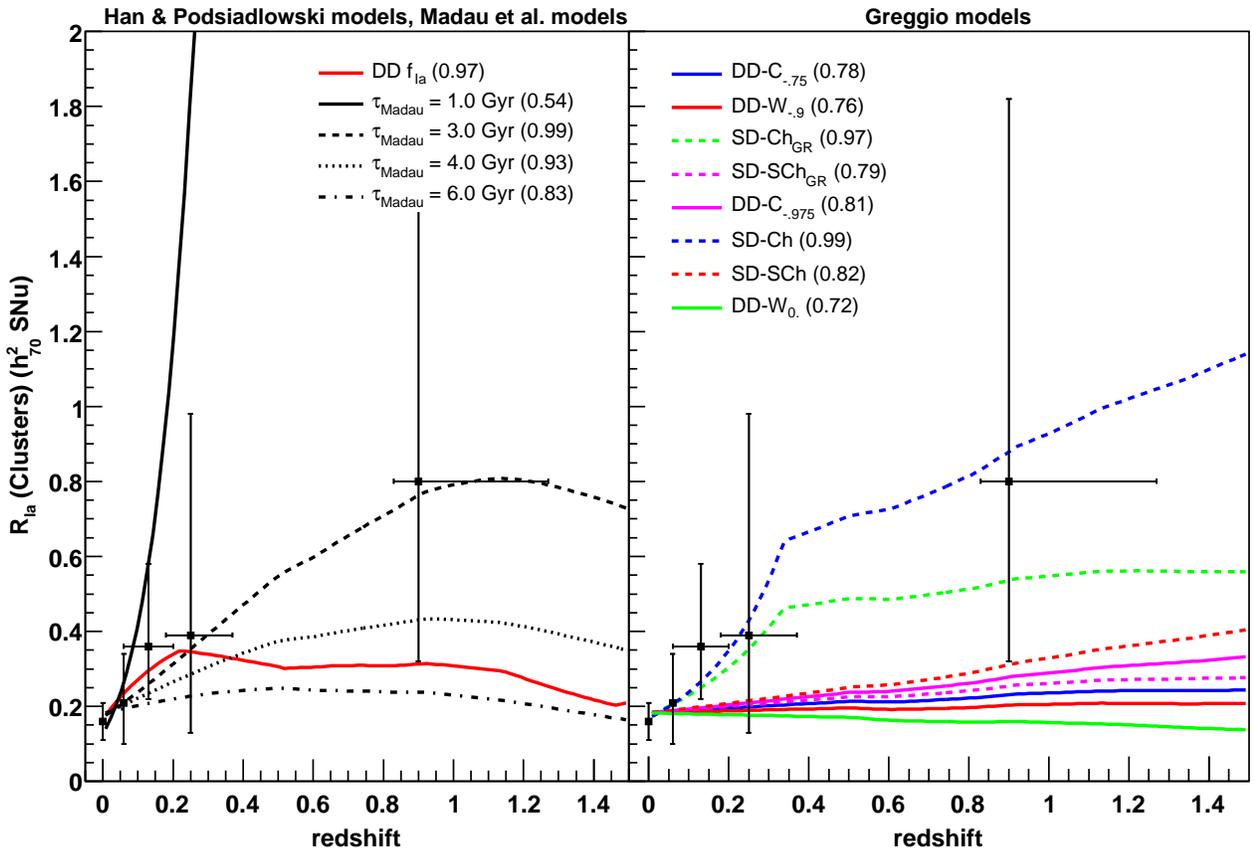}
\caption{SNIa rate measurements in clusters fitted with Han's and
  Madau's models (left), and Greggio's delay time distributions for an
  epoch of ellipticals formation of $z_f = 3$. $\chi^2$ fit
  probabilities are shown in parenthesis. Data are from
  \citet{cappellaro1999} at $z\sim 0$, \citet{reiss2000} at $z =0.11$,
  \citet{sharon2007} at $z=0.13$ and \cite{gal-yam2002} at $z=0.25$
  and $z=0.90$.}
\label{fig:clusterSNRIa_gre}
\end{figure*}

Both the individual rates and $f_{B,i}$ factors vary with redshift, so
that in order to compute the expected trend for the rate in clusters
one should specify the galaxy population in detail. On the other hand,
in spite of the observed variation of the cluster galaxy population
with redshift \citep{dressler1997,poggianti2006}, the evolution of the
bright portion of the luminosity function in clusters is consistent
with that of a passively evolving model in which galaxies form their
stars in a single burst at high ($z\sim 2,3$) redshift
\citep{depropris1999}. This justifies approximating the right hand
side of Eq.~(\ref{eqn:rIaclsnu}) with the evolution of the rate
obtained for such stellar population. By applying
Eq.~(\ref{eqn:rateIaburst}) we have:
\begin{equation}
\mathcal{R}_{\mathrm{Ia},\mathrm{SNu}}(t) \simeq 
k_{\Gamma} A_{\mathrm{Ia}} \cdot
(\mathcal{M}/L_{B}) \cdot f_{\mathrm{Ia}}(\tau=t-t_{f})
\label{eqn:rIaclsnuap}
\end{equation}
where $t$ is age of the universe at a given redshift, and $t_{f}$ is
the formation epoch of the stellar population.  Thus, the SNIa rate in
galaxy clusters as a function of redshift (in SNu) scales as the
product of the mass-to-light ratio and the distribution function of
the delay times, both of which depend on the age of the stellar
population at the various redshifts.  The fit of this relation to the
corresponding data yields a value for the $k_{\Gamma} A_{\mathrm{Ia}}$
factor.  \citet{sharon2007} present the SNIa rate in clusters as a
function of redshift. We plot their data in
Fig.~\ref{fig:clusterSNRIa_gre}, together with our fits\footnote{The
fit for Han SD model cannot be performed, since the SNIa rate from
this model is zero when the population is older than 1.5 Gyr (see
Fig.~\ref{fig:fIas}).} obtained with a formation redshift of 3. We
checked the dependence of the fit on formation redshift by adopting
$z_f = 2,4$ and found it to be small. We have included in the fit the
local value as measured by \citet{cappellaro1999} for E/SO galaxies in
the field.  Actually, even if ellipticals in the field may be slightly
younger than those in clusters \citep[e.g.][]{thomas2005}, the low
sensitivity of our fit to the formation redshift makes the use of this
local measurement reliable. If we exclude the \citet{cappellaro1999}
point from the fit, we derive a $\sim$ 40~\% larger value for both
$k_{\Gamma}A_{\mathrm{Ia}}$ and its error bar.  The mass-to-light
ratios as function of redshift have been derived by integrating
\citet{girardi2002} solar metallicity isochrones with a straight
Salpeter IMF.

Our fit with Madau's functions shows that the data favour
relatively large values of $\tau_{\mathrm{Madau}}$: the fit with 
$\tau_{\mathrm{Madau}}=1$ implies a too rapid growth of the SNIa rate with
redshift. This result is different from that in \citet{maoz2004}, who,
based on the same data, conclude for a relatively short
$\tau_{\mathrm{Madau}}$ ($< 2$ Gyr for a formation redshift of 2). The
different conclusion likely follows from the different approach; in
particular we consider the variation of the mass to light ratio with
redshift, while they do not. The decrease of $M/L_B$ with increasing
redshift counterbalances the growth of the distribution function of
the delay times, resulting into a slower expected evolution of the
rate in SNu with redshift.  

Fig.~\ref{fig:clusterSNRIa_gre} shows that due to the large error
bars, no real constraint on the distribution function of the delay
times can be derived at present. Most of the error bar reflects the
low number statistics \citep{sharon2007} so that future surveys will
greatly improve the situation. The models in
Fig.~\ref{fig:clusterSNRIa_gre} illustrate the diagnostic from this
kind of data: if the rapid growth of the SNIa rate with redshift is
confirmed, the DD models in Greggio's formulation would be unfavoured,
since they correspond to a flat behaviour of the rate at redshifts
smaller than $\sim 1$. The steep trend exhibited by the SD
Chandrasekhar models (blue and green dashed lines) mirrors the steep
late epoch decline of the distribution function of the delay times for
these kind of progenitors, due to the requirement that the WD reaches
the Chandrasekhar mass by accreting the envelope of a progressively
less massive companion \citep[see e.g.][ for more
details]{greggio2005}.

The derived values for the number of SNIa from 1~M$_\odot$ stellar
population are affected by a larger uncertainty with respect to what
we get from fitting the rate in the field. Averaging over all
Greggio's models we get $k_{\Gamma} A_{\mathrm{Ia}}=1.7\cdot 10^{-3}\
\mathrm{M}_{\odot}^{-1}$, not far from those obtained in the previous
section.  We acknowledge a difference of a factor of $\sim$ 3,
favouring events in clusters, at a $\sim$ 2 $\sigma$ level.  Given the
different ingredients used in the two computations, assumptions on the
global SFH and age distribution, and the uncertainty on the observed
rates, we regard this result as encouraging, although the discrepancy
needs further investigation.

An independent constraint on the $k_{\Gamma}A_{\mathrm{Ia}}$ factor
can be derived by considering the total Fe budget of galaxy clusters.
Assuming that every SNIa event ejects an average mass
$m_{\mathrm{Fe}}$ of Fe, their contribution to the Fe Mass-to-Light
ratio is:
\begin{equation}
\frac{M_{\mathrm{Fe,Ia}}}{L_B}=m_{\mathrm{Fe}} \cdot k_{\Gamma}
A_{\mathrm{Ia}} \cdot \frac{M}{L_B}
\end{equation}
where $M/L_B$ is the ratio between the total gas mass turned into
stars and the current $B$ band luminosity of the typical elliptical in
the cluster.  For a Salpeter IMF, $M/L_B \simeq 13.5\
\mathrm{M}_{\odot}\cdot \mathrm{L}_{\odot}^{-1}$ at an age of 11.4
Gyr, corresponding to a formation redshift of 3.  The total Fe mass to
light ratio in clusters is $\sim$ 0.02 \citep[e.g.][]{degrandi2004};
$\sim$ half of which is contributed from SNIa (see Sect. 5, also
Greggio 2008, in preparation).  Adopting $m_{\mathrm{Fe}}=0.75$
M$_\odot$ \citep[][ summing the ejected masses of $^{54}$Fe and
$^{56}$Fe]{iwamoto1999}, we get $k_{\Gamma}A_{\mathrm{Ia}} \sim
10^{-3}\ \mathrm{M}_{\odot}^{-1}$.  We conclude that the three
independent estimates of the $k_{\Gamma}A_{\mathrm{Ia}}$ agree within
a factor of 2. The uncertainties currently affecting each of them
prevents a more accurate determination.

\subsection{CCSN progenitor masses from CCSN rate}
\label{sec:imffromCCSN}

Due to the short lifetime of their progenitors, in a stellar system 
the CCSN rate is directly proportional to the current SFR
(Eq.~\ref{eqn:ccvstime}), with:
\begin{equation}
k_{\Gamma}A_{\mathrm{CC}} = 
\frac{\int_{m_{\mathrm{CC}_{\mathrm{min}}}}^{m_{\mathrm{CC}_{\mathrm{max}}}} 
  \Phi(m) dm}
{\int_{m_{{\star}_{\mathrm{min}}}}^{m_{{\star}_{\mathrm{max}}}}m\Phi(m) dm}
\label{eqn:scalingfactor}
\end{equation}
where $\Phi$ is the IMF by number, $[m_{{\star}_{\mathrm{min}}},
m_{{\star}_{\mathrm{max}}}]$ is the whole stellar mass range and
$m_{\mathrm{CC}_{\mathrm{min}}}$ and $m_{\mathrm{CC}_{\mathrm{max}}}$
are respectively the minimum and maximum stellar mass leading to a
CCSN. 

\begin{figure*}[!htb]
\centering
\includegraphics[width=17cm]{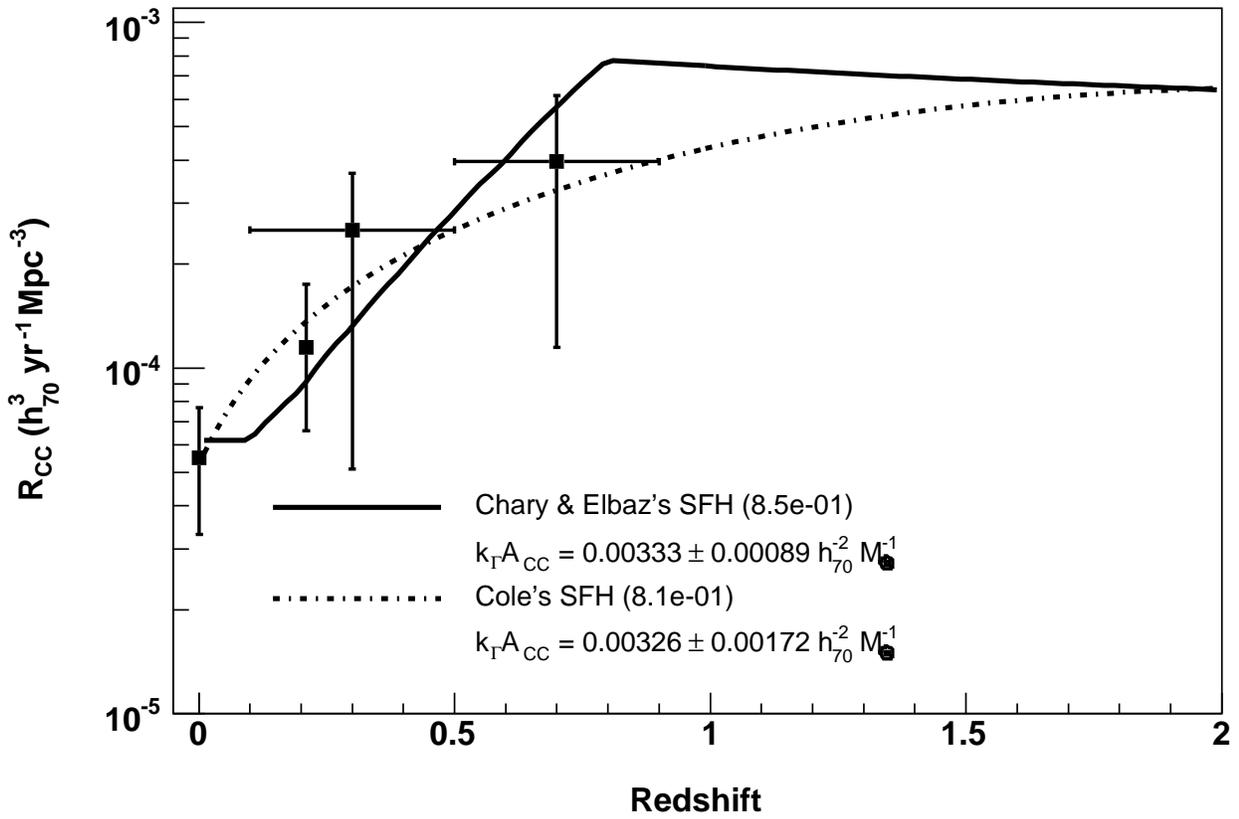}
\caption{Scaling of the SFH to the CCSN rate. Data are from
  \citet{cappellaro1999} at $z \sim 0$, \citet{botticella2008} at $z =
  0.30$ and \citet{dahlen2004} at $z=0.3$ and $z=0.7$ (see
  Tab.~\ref{tab:otheresultsCC}). The fitted value of $k_{\Gamma}
  A_{\mathrm{CC}}$ is shown; the $\chi^2$ probability is in
  parenthesis.}
\label{fig:snr2toimf}
\end{figure*}

\begin{figure*}[!htb]
\centering
\includegraphics[width=17cm]{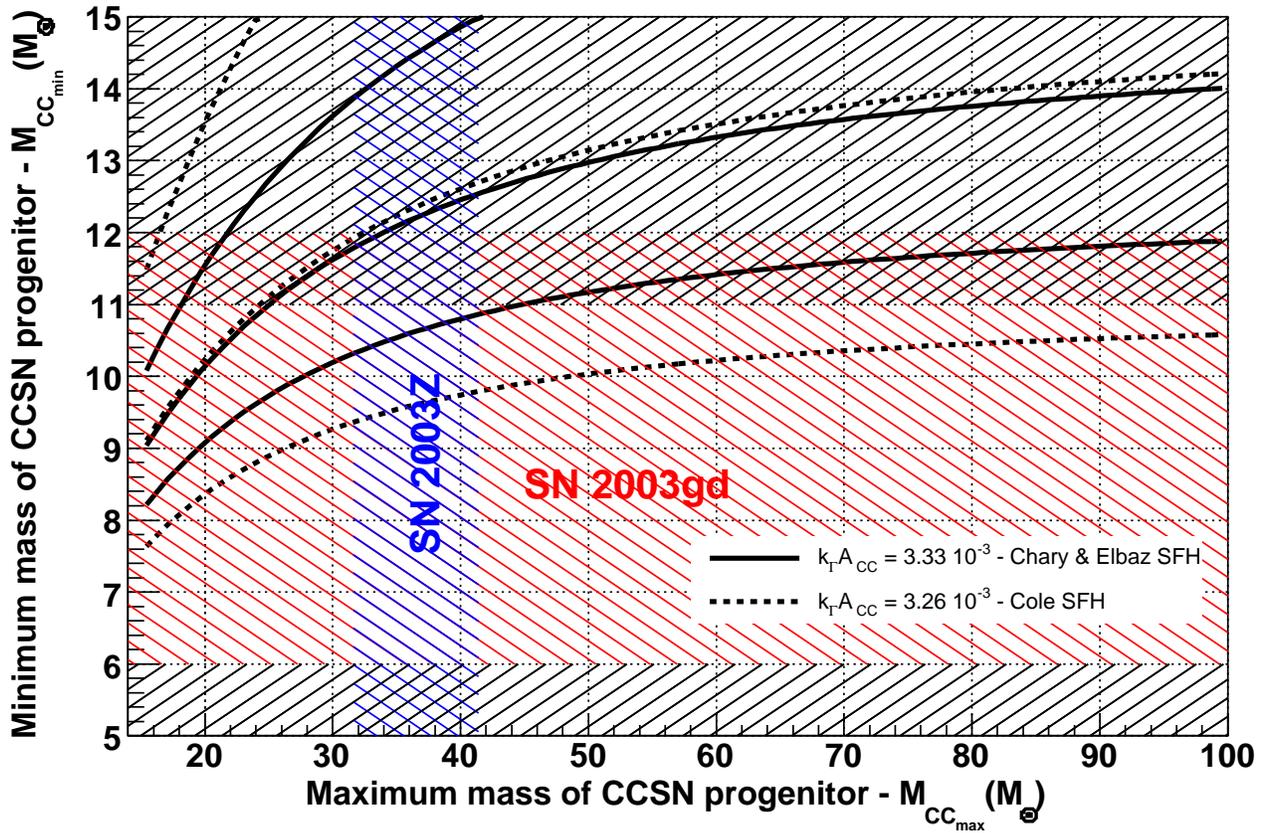}
\caption{Minimal CCSN progenitor mass as a function of the maximum
  mass, for two values of the scaling factor
  ($k_{\Gamma}A_{\mathrm{CC}}$) between CCSN rate and SFH, with their
  1-$\sigma$ error (the lines from top to bottom represent
  $k_{\Gamma}A_{\mathrm{CC}} - \sigma$, $k_{\Gamma}A_{\mathrm{CC}}$
  and $k_{\Gamma}A_{\mathrm{CC}} + \sigma$ for both plotted values;
  $k_{\Gamma}A_{\mathrm{CC}}$ increases from top to bottom). Hatched
  zones (black parallel lines) reflect the theoretical ``forbidden''
  area for the minimum CCSN progenitor mass, according to
  \citet{heger2003}; red and blue hatched areas stand respectively for
  SN 2003gd and SN 2003Z progenitor mass intervals.}
\label{fig:min_mass_max_mass}
\end{figure*}

For a Salpeter IMF between 0.1 and 120~M$_\odot$:
\begin{equation}
k_{\Gamma}A_{\mathrm{CC}} = 0.126 \cdot 
(m_{\mathrm{CC}_{\mathrm{min}}}^{-1.35} -
  m_{\mathrm{CC}_{\mathrm{max}}}^{-1.35}).
\label{eqn:scalingfactor1}
\end{equation}
By fitting Eq.~(\ref{eqn:ccvstime}) to the redshift trend of the CCSN
rate with the two SFH laws described in Sec. \ref{sec:sfh} we obtain
the related values of the constant $k_{\Gamma} A_{\mathrm CC}$.
Although the scaling of the SFH to the CCSN rate depends on the IMF,
we only consider Salpeter IMF because this has been adopted for the
derivation of the SFH laws. The fitting values of $k_{\Gamma}
A_{\mathrm CC}$ are reported in Fig.~\ref{fig:snr2toimf}, giving
similar values for CE and Cole SFH.  For a given SFH we get a
bijective relationship between $m_{\mathrm{CC}_{\mathrm{min}}}$ and
$m_{\mathrm{CC}_{\mathrm{max}}}$, which is very sensitive to
$k_{\Gamma} A_{\mathrm CC}$, as plotted in
Fig.~\ref{fig:min_mass_max_mass}. 

The minimum mass $m_{\mathrm{CC}_{\mathrm{min}}}$ is actually set by
the heaviest star that produces a white dwarf, which can vary from $6\
\mathrm{M}_{\odot}$ to $11\ \mathrm{M}_{\odot}$ according to
\citet{heger2003}. They used $9\ \mathrm{M}_{\odot}$ as an average
value.  It is supported by the observations of SNII events: the lowest
SNII progenitor mass known so far has been derived from the type II-P
event SN~2003gd as $8^{+4}_{-2}\ \mathrm{M}_{\odot}$
\citep{smartt2004,hendry2005}, though careful modeling of the
observations favors a progenitor mass closer to the upper limit of
this range \citep{hendry2005}. The maximum mass
$m_{\mathrm{CC}_{\mathrm{max}}}$ is much less constrained, but should
be around $\sim 40\ \mathrm{M}_{\odot}$, or less. The highest main
sequence masses that have been derived for CCSN event are for some
type Ic connected with a hypernova event, with
$m_{\mathrm{CC}_{\mathrm{max}}} \gtrsim 40\ \mathrm{M}_{\odot}$. This
kind of SN is of course included in the generic class of ``CCSN
event'' for which the rate history has been measured.  For instance
the type Ic SN~2003lw has a progenitor with a mass within $[40-50\
\mathrm{M}_{\odot}]$ \citep{mazzali2006}. Within the type II-P SNe
class, SN~2003Z has the highest ``observed'' mass with
$m_{\mathrm{CC}_{\mathrm{max}}} = 36.5\pm 4.9\ \mathrm{M}_{\odot}$
\citep{pastorello2003,zampieri2003}.

Fig.~\ref{fig:min_mass_max_mass} shows that the observed and
theoretical $m_{\mathrm{CC}_{\mathrm{min}}}$ and
$m_{\mathrm{CC}_{\mathrm{max}}}$ are only compatible with the high
side of the 1-$\sigma$ error interval of $k_{\Gamma} A_{\mathrm CC}$
for both CE and Cole SFH. The favored $m_{\mathrm{CC}_{\mathrm{min}}}$
is $m_{\mathrm{CC}_{\mathrm{min}}} \gtrsim 10\ \mathrm{M}_{\odot}$ to
account for the observed $m_{\mathrm{CC}_{\mathrm{max}}}$. 
But since CCSNe explode in star forming regions, a fraction of the
events is likely to escape detection, so that the rate in
Fig.~\ref{fig:snr2toimf} would be a lower limit to the real CCSN rate
(see sec.~\ref{sec:measurements}). In this case, the fit with the SFH
would lead to a larger value for $k_{\Gamma}A_{\mathrm{CC}}$, which
could accommodate a wider range of CCSN progenitor's masses.

\section{The released iron density evolution}
\label{sec:iron}

The iron in the Universe is produced or released in the ISM by
supernovae. Knowing the supernova rate history and the distribution of
produced iron, we can estimate the evolution of the iron density with
redshift. As a function of cosmic time, $t$, it is generally given by:
\begin{equation}
\rho_{\mathrm{Fe}}(t) = \int_{t_{\mathrm{re}}}^{t}
\mathcal{M}_{\mathrm{Fe}}(t')\cdot \mathcal{R}_{\mathrm{SN}}(t') dt'
\label{eqn:FeSNe}
\end{equation}
where $\mathcal{M}_{\mathrm{Fe}}$ is the distribution function of iron
mass released per event and $t_{\mathrm{re}}$ is the epoch of
reionization corresponding to the first stars, and then the first SNe,
and first iron released in the ISM.

\subsection{From type Ia SNe}
\label{sec:ironIa}

A SNIa releases in the interstellar medium (ISM), on average,
0.5~M$_{\odot}$ of $^{56}$Ni (that becomes of $^{56}$Fe after
radioactive decay). During the explosion others isotopes of iron are
produced, $^{54}$Fe, $^{57}$Fe and $^{58}$Fe. The present approach is
sensitive only to the ''observable'' iron which powers the light curve
of SNIa, that is $^{56}$Fe.  Then we expect to miss 10~\% to 15~\% of
iron according to delayed detonation nucleosynthesis models
\citep{iwamoto1999}; for 3-dimensional pure deflagration models, the
proportion of iron isotopes other than $^{56}$Fe varies from 15~\% to
25~\% \citep{travaglio2004}. To take into account all iron, we assume
that $^{56}$Fe represents 85~\% of all iron released by SNIa in the
ISM. Then, in Eq.~(\ref{eqn:FeSNe}) the iron distribution is given by
$\mathcal{M}_{\mathrm{Fe}} = 1.18\ \mathcal{M}_{^{56}\mathrm{Fe}} =
1.18\ \mathcal{M}_{^{56}\mathrm{Ni}}$, assuming that this ratio does
not evolves with cosmic time\footnote{Actually this ratio probably
  evolves with redshift since \citet{tavaglio2005} found an
  anticorrelation of
  $\mathcal{M}_{^{56}\mathrm{Fe}}/\mathcal{M}_{\mathrm{Fe}}$ with
  metallicity, going from 0.93 at $0.1\ \mathrm{Z}_{\odot}$, to 0.83
  at $\mathrm{Z}_{\odot}$ and 0.70 at $3\ \mathrm{Z}_{\odot}$. But
  such a study is beyond the scope of this paper.}.

\begin{figure*}[!htb]
\centering
\includegraphics[width=17cm]{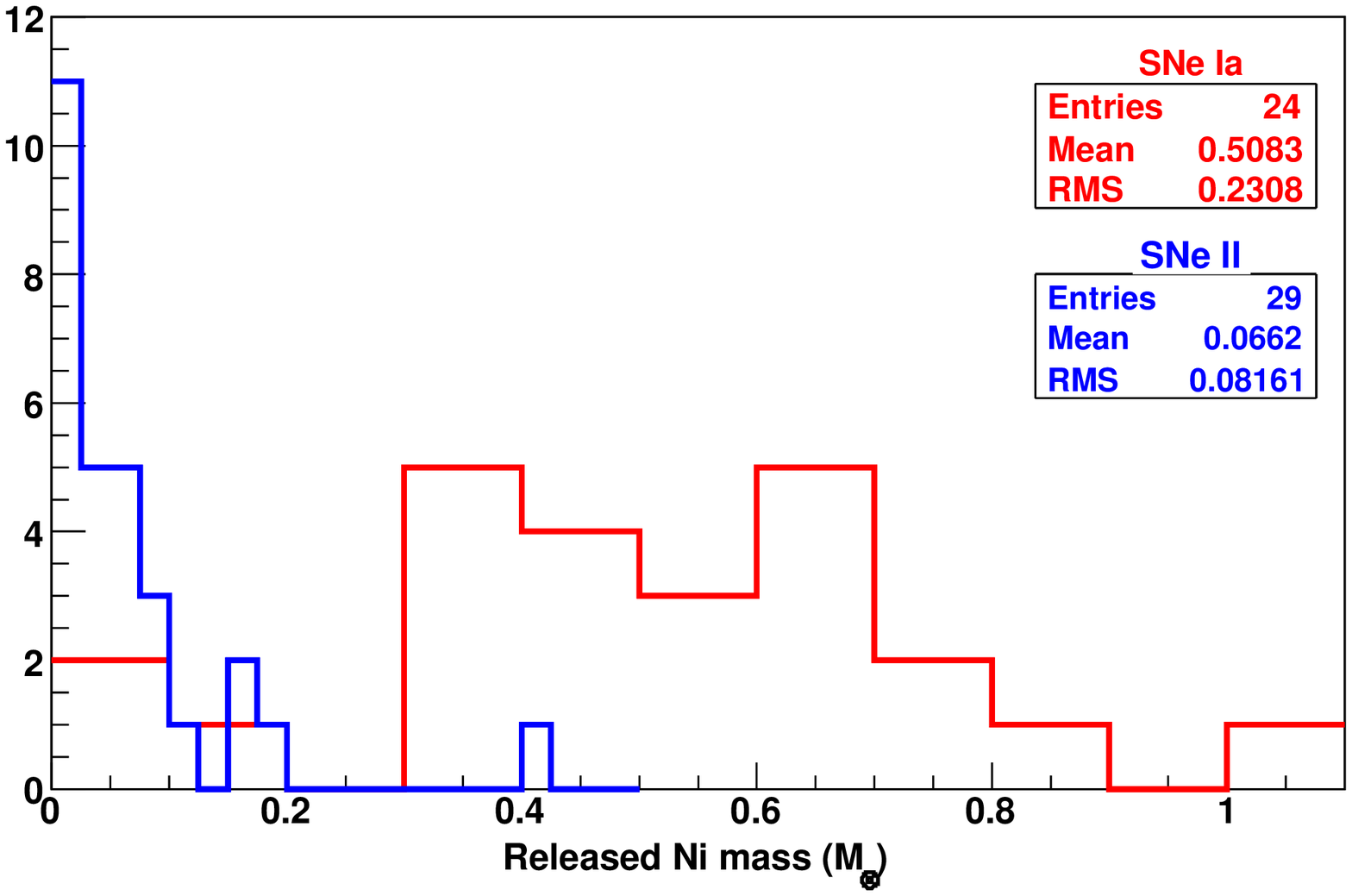}
\caption{Histogram of released $^{56}$Ni mass,
 $\mathcal{M}_{^{56}\mathrm{Ni}}$, for type Ia SNe
 \citep{leibundgut2000,mazzali2000,stritzinger2006}; we took the mean
 value when necessary. And for type II SNe (\citet{zampieri2003} and
 \citet{hamuy2003}; we took the \citet{zampieri2003} values for
 identical SNe in both sample).}
\label{fig:histomNiSNe}
\end{figure*}

The distribution of released $^{56}$Ni mass is taken from
\citet{leibundgut2000}, \citet{mazzali2000} and
\citet{stritzinger2006}; it is plotted on
Fig.~\ref{fig:histomNiSNe}. The averaged $^{56}$Ni mass as derived
from observations, $0.5\ \mathrm{M}_{\odot}$ per event, lies between
the prediction of DD models ($0.6\ \mathrm{M}_{\odot}$ to $0.8\
\mathrm{M}_{\odot}$, \citet{iwamoto1999}) and predictions of pure
3-dimensions deflagration models ($0.24\ \mathrm{M}_{\odot}$ to $0.44\
\mathrm{M}_{\odot}$, \citet{travaglio2004}).

\begin{figure*}[!htb]
\centering
\includegraphics[width=17cm]{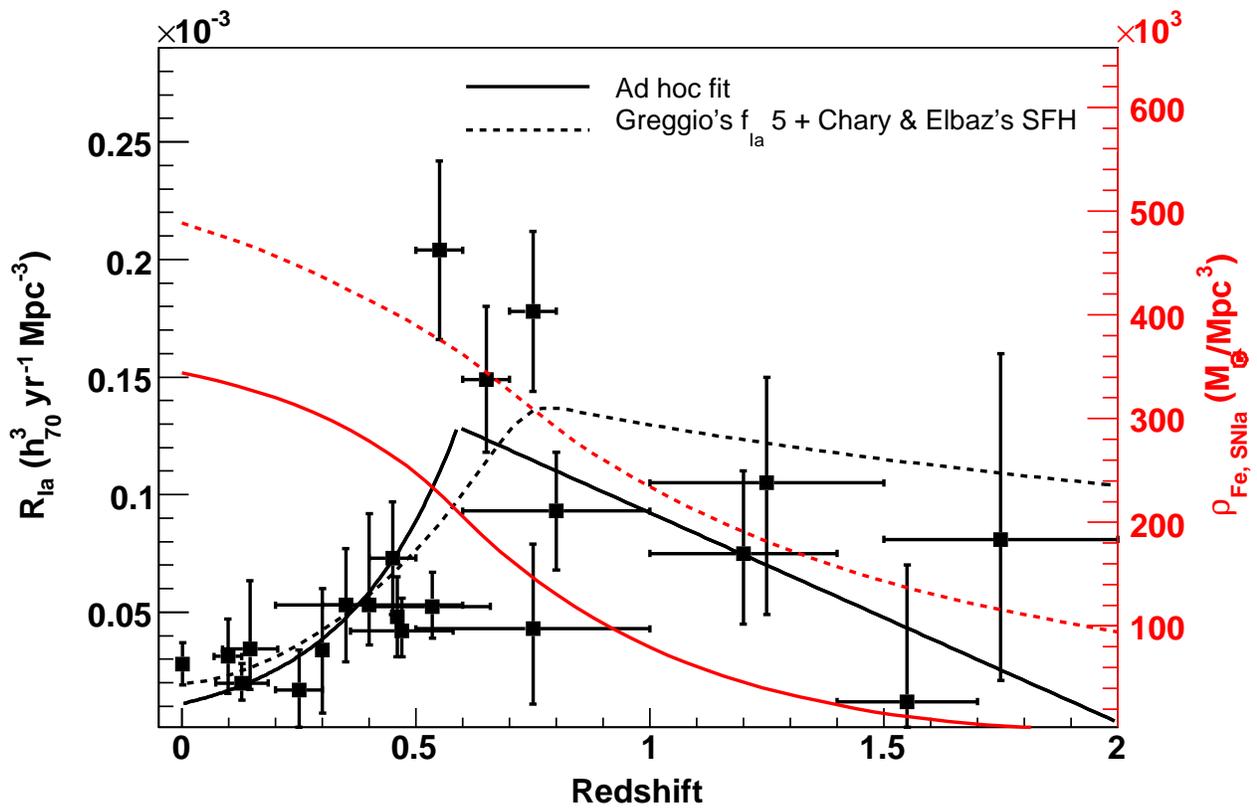}
\caption{Density of iron released by SNe Ia (in red, right scale) for
  two models of SN rate evolution. The SNIa rate measurements are
  superimposed (left scale) and fitted by an exponential between $z =
  0$ and 0.59, and by a straight line for $z > 0.59$, or modeled by
  the Greggio $\mathrm{SD}_{\mathrm{subCh}}$ delay time distribution
  convolved with the CE SFH (in black, left scale).}
\label{fig:FeSNeIa}
\end{figure*}

To render the SNIa rate measurement and integrate it easily, we fit it
by an exponential between $z=0$ and 0.59, and by a straight line for
$z > 0.59$ as shown on Fig.~\ref{fig:FeSNeIa}. Eq.~(\ref{eqn:FeSNe})
is integrated by a Monte-Carlo method in order to keep the observed
shape of the nickel distribution, under the assumption that the
distribution of $^{56}$Ni as shown on Fig.~\ref{fig:histomNiSNe} is
representative and independent of cosmic time. Beside the \textit{ad
hoc} shape for the SNIa rate, we use the best model fitting the rate
evolution, that is the Greggio SD-SCh distribution convolved with the
CE SFH (Fig.~\ref{fig:SNRIa_z_Greggio}). The result is plotted on
Fig.~\ref{fig:FeSNeIa}.

Since the model produces SNe Ia at redshift larger than $\sim 2$, for
this case we get a larger amount of integrated iron, with respect to
the \textit{ad hoc} fit. Still, the difference only amounts to a
factor of 1.2 at the present epoch. So SNe Ia in the distant universe
have little influence on the total iron produced.

\subsection{From CCSNe}

Like for SNIa, only $^{56}$Ni powers the light curve of CCSN, by
radioactive decay to $^{56}$Fe. According to \citet{iwamoto1999} a
typical CCSN ejects 93 \% of the total iron as $^{56}$Fe; thus in our
total computation, we take $\mathcal{M}_{\mathrm{Fe}} = 1.08\
\mathcal{M}_{^{56}\mathrm{Fe}} = 1.08\ \mathcal{M}_{^{56}\mathrm{Ni}}$
in Eq.~(\ref{eqn:FeSNe}).

\begin{figure*}[!htb]
\centering
\includegraphics[width=17cm]{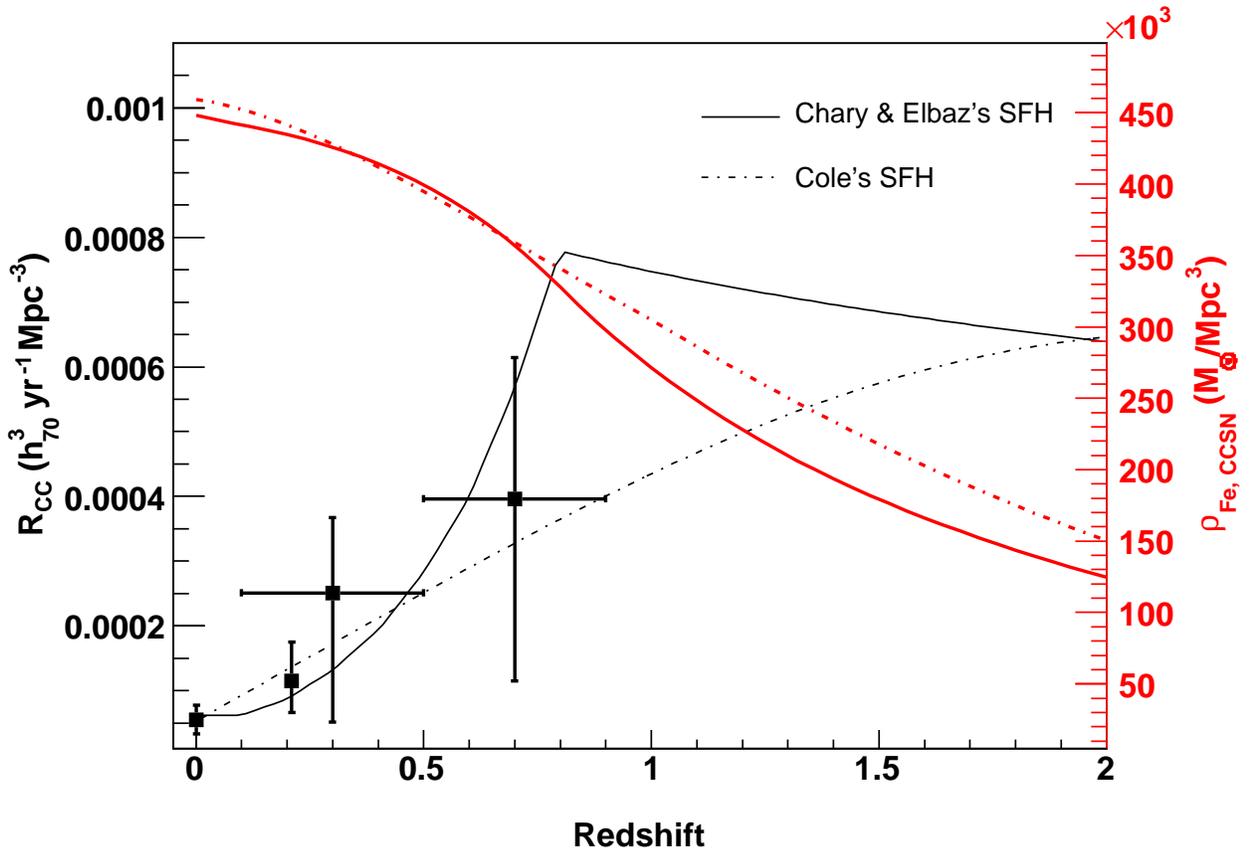}
\caption{Density of iron released by CCSNe (in red, right scale) as
  computed using different SFR models to fit the CCSN rate (in black,
  left scale). The first stars formation redshift is set to 20.}
\label{fig:FeSNeCC}
\end{figure*}

According to \citet{zampieri2003} and \citet{hamuy2003} the $^{56}$Ni
mass released by SNII is distributed as shown in
Fig.~\ref{fig:histomNiSNe}, based on 29 SNe II. The average Ni mass is
0.066 $\mathrm{M}_{\odot}$. This distribution covers more than two
orders of magnitudes, while the distribution of Ni produced by SNIa
spreads less than one order of magnitude.  Eq.~(\ref{eqn:FeSNe}) is
integrated over this distribution by a Monte-Carlo method, setting the
redshift of the first stars to 20 \citep{fuller2000}. The CCSN rate is
modeled by two SFH models as previously discussed. They are plotted
on Fig.~\ref{fig:FeSNeCC}, together with the results of the
integration.  The two SFH models, once calibrated on the observed
rates, produce very close Fe production curves.

\subsection{Total}

\begin{figure*}[!htb]
\centering
\includegraphics[width=17cm]{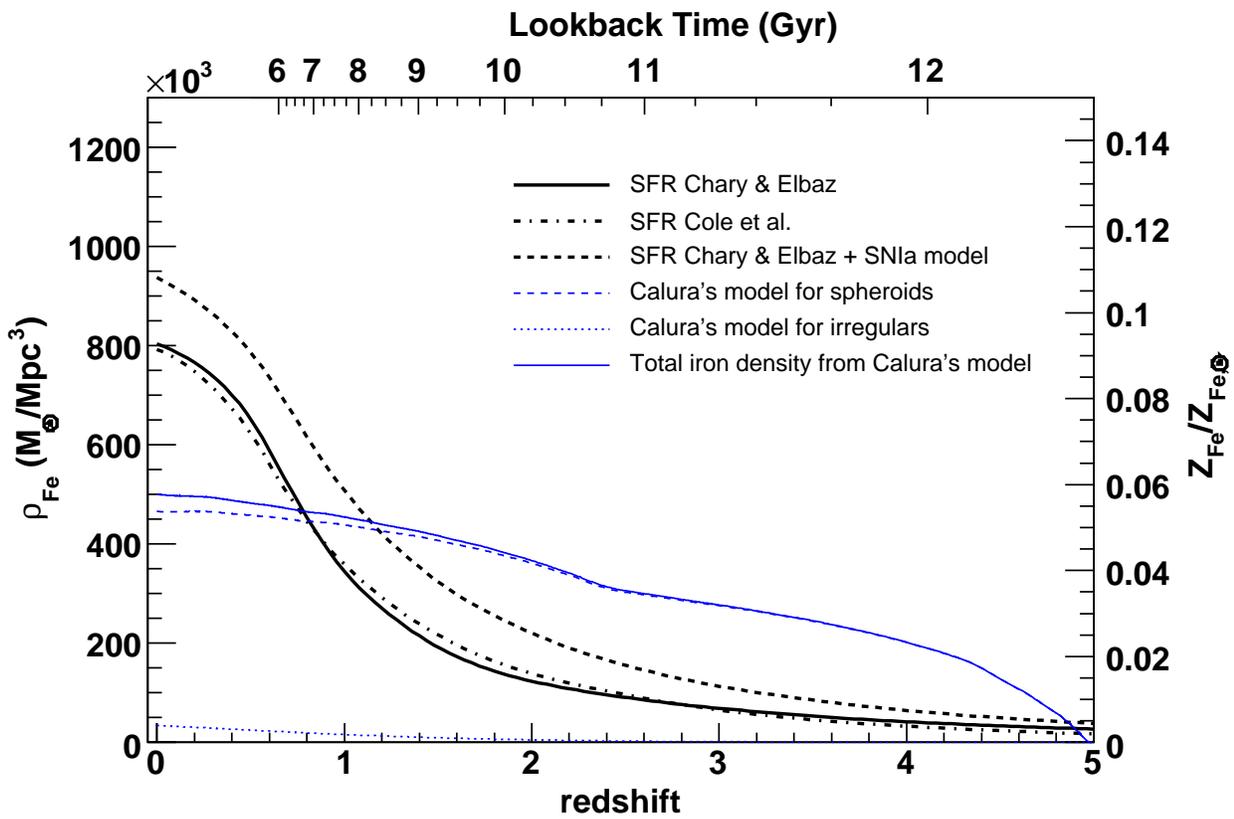}
\caption{Density of Fe released by both type Ia and CCSNe (for
  different SFH models) as a function of redshift (in black). The
  epoch of the first star explosions is set to $z = 20$. Right axis
  shows the corresponding iron mass abundance. Also shown (in blue) is
  the model by \citet{calura2006} of produced iron evolution, for
  dwarfs and irregular galaxies (dots), for spheroids (dashes) and in
  total (solid).}
\label{fig:IronSNe}
\end{figure*}

Fig.~\ref{fig:IronSNe} shows the total density of iron produced, sum
of the contribution of SNIa and of CCSN.  It is interesting to note
that today's iron density comes from half-part of SNIa
(Fig.~\ref{fig:FeSNeIa}) and half-part of CCSN
(Fig.~\ref{fig:FeSNeCC}). Indeed, on average one CCSN releases
10 times less Fe than one SNIa. However, the CCSN production per unit
gas mass transformed into stars is about 10 times that of SNIa
production, as result from the ratio $k_{\Gamma}
A_{\mathrm{CC}}/k_{\Gamma} A_{\mathrm{Ia}} \sim 7.5$.  We remark that
the evolution of the iron density does not depend strongly on the SFH
model.  \citet{calura2006} present models of cosmic metal production
based on chemo-photometric models of the evolution of dwarf galaxies
and spheroids in the universe. The models detail the evolution of
various chemical species in different phases (ISM, IGM, Stars), and,
in particular, that of the total Fe density. Fig.~\ref{fig:IronSNe}
shows as blue lines the predictions of Calura \& Matteucci models for
two kinds of galaxies, irregulars and spheroids, and their sum. Since
the contribution of disks is missing, we cannot compare directly the
Calura and Matteucci predictions (solid blue line) with our empirical
curve.  However, in \citet{calura2004} the iron production rate by
disks becomes important only at relatively low redshift ($z \lesssim
1.2$), and at zero redshift the contribution of disks is $\sim$ 0.6
that of ellipticals. We may then extrapolate the results of
\citet{calura2006} into a total expected Fe production in the Universe
of $8 \cdot 10^5\ \mathrm{M}_{\odot}\ \mathrm{Mpc}^{-3}$. This is in
excellent agreement with our estimate, which is $\rho_\mathrm{Fe} =
(8.5\pm 0.9)\cdot10^5\ \mathrm{M}_{\odot}\ \mathrm{Mpc}^{-3}$, as
averaged from our various estimates. We also expect that the corrected
curve for the total Fe in Calura and Matteucci's models is similar to
what plotted in Fig.~\ref{fig:IronSNe} for redshift in excess of
$\sim$~1.2, so that also in the models a large fraction of Fe in the
universe is produced at $z \lesssim 1$.  Given the completely
different approaches for determining the cosmic increase of total Fe
density, we consider the models and our empirical estimate in very
good agreement.  Our total current Fe density is also in excellent
agreement with the value of $\rho_\mathrm{Fe} = 8.6\cdot 10^5\
\mathrm{M}_{\odot}\ \mathrm{Mpc}^{-3}$ by \citet{fukugita2004}.

We label the right axis in Fig.~\ref{fig:IronSNe} with the Fe
abundance relative to solar, having considered $Z_\mathrm{Fe} =
\rho_\mathrm{Fe}/\rho_B$, where $\rho_B = \rho_{\circ}\cdot \Omega_B$,
$\rho_{\circ} = 1.36\cdot 10^{11}\ h_{70}^2 \mathrm{M}_{\odot}
\mathrm{Mpc}^{-3}$ is the critical density of the universe and
$\Omega_B$ is the density of baryon \citep[$\Omega_B = 0.049\pm 0.002\
h_{70}^{-2}$, see ][]{spergel2003}, which gives: $Z_{\mathrm{Fe}} =
(1.28 \pm 0.14)\cdot 10^{-4}$. Adopting the solar Fe abundance by
\citet{grevesse1998} $Z_{\mathrm{Fe},\odot} = (1.3 \pm 0.1) \cdot
10^{-3}$, we obtain an average Fe abundance of the universe at zero
redshift of $Z_{\mathrm{Fe}} = (0.10\pm 0.01)\ Z_{\mathrm{Fe},\odot}$,
which is a factor of $\sim$~3-4 lower than the Fe abundance in
clusters of galaxies \citep[e.g.][]{renzini2004}. 

Does this mean that the stellar production of Fe is different in
clusters from that in the field? To answer this question we consider
the ratio between the Fe and the stellar density $\eta_{\mathrm{Fe}} =
\rho_{\mathrm{Fe}}/\rho_*$. In clusters: $\eta_{\mathrm{Fe,cl}} =
Z_{\mathrm{Fe,ICM}}\cdot M_{\mathrm{ICM}}/M_* + Z_{\mathrm{Fe},*}
\simeq 3.3\ Z_{\mathrm{Fe},\odot}$ having adopted
$M_*/M_{\mathrm{ICM}}=0.13$ \citep{ettori2007},
$Z_{\mathrm{Fe,ICM}}=0.3 Z_\odot$ \citep{tamura2004}, and a solar iron
abundance for the stellar component. By integrating our SFH and
correcting for the mass return from SNe and stellar winds\footnote{
for a Salpeter IMF the correction factor amounts to 0.72
\citep[e.g.][]{cole2001}}, we find a current $\rho_* = (8.2 \pm
4.1)\cdot 10^8\ h_{70}\mathrm{M}_{\odot}\cdot \mathrm{Mpc}^{-3}$,
which gives $\eta_{\mathrm{Fe}} = (0.8 \pm 0.4)\cdot \
Z_{\mathrm{Fe},\odot}$ a factor $\sim 4$ lower than in clusters.
However, while these estimates of $\eta_{\mathrm{Fe,cl}}$ and of the
Fe density in the field do not depend on the IMF, the stellar density
in the field does.  A more realistic IMF, with a turn-over at the low
mass end, implies a lower stellar density for the same amount of UV
photons. For example, the mass fraction in stars more massive than 1
M$_{\odot}$ is 0.61 and 0.39 for Chabrier \citep{chabrier2005} and
Salpeter IMFs respectively; the mass fraction returned to the medium
from stellar winds and SNe is $\sim$~0.45 and 0.28 for the two
IMFs. Therefore, for a Chabrier IMF, the current stellar mass density
becomes $\rho_*^{\mathrm{Cha}} = (4 \pm 2)\cdot 10^8\
\mathrm{M}_{\odot} \mathrm{Mpc}^{-3}$, to yield
$\eta_{\mathrm{Fe}^{\mathrm{Cha}}} = (1.6 \pm 0.8)\
Z_{\mathrm{Fe},\odot}$, close to the value in clusters. In addition,
the rate of CCSN may be affected by a more severe incompleteness than
adopted in the values in Tab.~\ref{tab:otheresultsCC}, particularly at
high redshift, where some events could occur in heavily obscured
regions \citep{mannucci2003}.  We conclude that the Fe production from
stars in the field and in clusters can be brought into agreement, if
the IMF has a turn over at the low mass end.  The different abundance,
then, stems from a different efficiency of baryon conversion into
stars $\rho_*^{\mathrm{Cha}}/\rho_B$, which adopting the figures
above, is $\sim 0.12$ in clusters and $\sim 0.06$ in the field.

\section{Conclusions}
\label{sec:conclusions}

In this paper we have investigated on the constraints which can be
derived from the cosmic trend of the SN rates on the SFH, on the
distribution of the delay times of SNIa progenitors and on the mass
range on the CCSN progenitors. In addition, we have determined the
local Fe abundance as resulting from the integration of the observed
SN rates. Our results can be summarized as follows.

The observed cosmic evolution of the SNIa rate has been fit by
adopting two laws for the SFH which bracket most of the observational
data, namely \citet{chary2001} and \citet{cole2001}, in combination
with various distribution functions of the delay times:
\citet{madau1998} models, a selection of \citet{greggio2005} models,
and a selection of models by \citet{han2003,han2004}. We find that,
while, within the error, it is hard to distinguish among the possible
progenitor models, the fits are fairly sensitive to the adopted SFH
\citep[see also][]{oda2008}. The increase of the SNIa rate with
redshift is better reproduced with CE rather than Cole SFH. However if
rate measurements by \citet{barris2006} were systematically too high,
the two SFH would be rather equivalent in fitting the SNIa rate data.
The inability of the cosmic SNIa rate to constrain the progenitor
model is related to its being an average over a galaxy population
which spans a wide range of properties. Measurements of the SNIa rate
in different kinds of galaxies would help us to relate the observed
rate, and its redshift evolution, to the parent stellar population, in
turn providing better constraints on the shape of the distribution
function of the delay times.
 
As an attempt in this direction, we have considered the redshift
evolution of the SNIa rate in galaxy clusters, assumed to harbor
mainly old stellar populations. Due to the large error bars affecting
the current measurements, all SNIa progenitor models provide
acceptable fits.  Madau's models with $\tau_{\mathrm{Madau}}$ shorter
than $\sim$ 1 Gyr are however excluded. Among Greggio's models, the SD
family seems favoured, but we emphasize that the error bars on the
data are currently too large to draw any conclusion.

The fit to the cosmic SNIa rate and to the rate in clusters provides
the number of Ia events per unit mass of the parent stellar population
($k_{\Gamma}A_{\mathrm{Ia}}$) which is needed to explain the observed
level of the rate. The field cosmic rate and the rate in clusters
indicate similar (albeit non equal) values for this constant. On the
average we conclude that in both field and cluster galaxies we have
$\sim$ 1 SNIa explosion every 1000 M$_{\odot}$ of stars in the parent
population, corresponding to $\sim$ a 5 \% probability of the SNIa
channel, if the primaries in the binary progenitors have
masses in the range 3 to 9 $\mathrm{M}_\odot$ (for a Salpeter IMF).

The redshift evolution of the rate of CCSN is compatible with both CE
and Cole SFH. The favored progenitor minimum mass is $\gtrsim 10\
\mathrm{M}_{\odot}$, even if some incompleteness in the observed CCSN
rate could accommodate a (somewhat) lower CCSN's progenitor masses.

The convolution of the observed cosmic SN rates with their ejected Fe
mass allows us to compute the Fe density of the universe at zero
redshift.  We find $\rho_\mathrm{Fe} = (8.8\pm 0.9)\cdot10^5\
\mathrm{M}_{\odot}\ \mathrm{Mpc}^{-3}$, in very good agreement with
\citet{fukugita2004} and \citet{calura2006}. The corresponding Fe
abundance in the local universe is $0.1\ Z_{\mathrm{Fe},\odot}$,
which is a factor of $\sim 3$ lower than the Fe abundance in massive
clusters. 
We show that, if the IMF has a turn over at the low mass end,
this difference does not imply a different efficiency of Fe production
from stars in the two environments, but rather it reflects a different
efficiency of baryon conversion to stars.

\section*{Acknowledgments}

We thank Andrew Hopkins for kindly providing his updated compilation
of SFR measurements; Philipp Podsiadlowski and Zhanwen Han for useful
discussions and for providing ascii forms of their SNIa delay time
distribution models; Francesco Calura for quickly providing of his
iron density evolution model in ascii form; Maria Teresa Botticella
for providing her results on the SN rate before publication. We also
thank Alvio Renzini and Enrico Cappellaro for interesting discussions.
This work has been partly supported by the European Community's Human
Potential Program under contract HPRN-CT-2002-00303, \textit{The
Physics of Type Ia Supernovae}.






\bibliographystyle{aa} 
\bibliography{biblio_na}

%
%

%
%


%
%


%
%
 

%
%


%
%

%

%
%


%
%


%
%


%
%


%
%


%
%


%
\renewcommand{\multirowsetup}{\centering}
\newlength{\LL}
%

\begin{table*}[!ht]
\centering
\begin{tabular}{lllcccccl}
\hline
      & \multicolumn{2}{c}{SNIa rate$^{\ a}$} 
&  & SNe &  survey& limiting & spectro & \\
\cline{2-3}
\raisebox{2.5ex}[0cm][0cm]{$\langle z\rangle$}  & ($h_{70}^2$ SNu) & 
($10^{-5}\ h_{70}^3\ \mathrm{Mpc}^{-3}\ \mathrm{yr}^{-1}$) & 
\raisebox{2.5ex}[0cm][0cm]{($\Omega_{M_{\circ}}, \Omega_{\Lambda_{\circ}}$)} & 
nb &
surface &
mag &
confirmed &
\multicolumn{1}{c}{\raisebox{2.5ex}[0cm][0cm]{author}}
\\ 
\hline
0.38  & $0.40^{+0.26+0.18}_{-0.18-0.12}$ & \multicolumn{1}{c}{-} & (1.0, 0.0) & 3 
& 1.7 deg$^2$ & $R \sim 23$ & 100 \% & \citet{pain1996} 
\\ 
0.01    & $0.18\pm0.05$ & 2.8$\pm$0.9 & & 70 
& $10^4$ gal. & $B \sim 20$ & 100 \% & \citet{cappellaro1999}$^b$ 
\\
0.14 & $0.22^{+0.17+0.06}_{-0.10-0.03}$ &
$3.43^{+2.7+1.1}_{-1.6-0.6} $& (0.3, 0.0) & 4 
& 80 deg$^2$ & $V = 21.5 $ & 100 \% & \citet{hardin2000}$^{b, c}$ 
\\
0.55 & $0.28^{+0.05+0.05}_{-0.04-0.04}$ &
  $5.25^{+0.96+1.10}_{-0.86-1.06}$ & (0.3, 0.7) & 38 
& \settowidth{\LL}{12 deg$^2$}
\multirow{2}{\LL}{12 deg$^2$}&
\settowidth{\LL}{$R \sim 24.5$}
\multirow{2}{\LL}{$R \sim 24.5$}& 
\settowidth{\LL}{100 \%}
\multirow{2}{\LL}{100 \%} & \citet{pain2002}  
\\ 
0.55 & $0.46^{+0.08+0.07}_{-0.07-0.0.7}$ &
$11.1^{+2.0+1.1}_{-1.8-1.1}$ & (1.0, 0.0) & 38 
& & & & \citet{pain2002}  
\\ 
0.098 & 0.196 $\pm$ 0.098 & $3.12 \pm 1.58$ &  & 19 
& $10^5$ gal& $V > 20.4$ & 100 \% & \citet{madgwick2003}$^b$  
\\  
0.46  & \multicolumn{1}{c}{-} & $4.8\pm 1.7$ & (0.3, 0.7) & 8 
& 2.5 deg$^2$ & $I \sim 24$& 100 \% & \citet{tonry2003}  
\\ 
0.13 & $0.125^{+0.044+0.028}_{-0.034-0.028}$  &
$1.99^{+0.70+0.47}_{-0.54-0.47}$  & (0.3, 0.7) &
 14 
& 598 deg$^2$ & $V \sim 21$ & 100 \% &  \citet{blanc2004}  
\\ 
$[0.2, 0.6]$ & \multicolumn{1}{c}{-} & $6.9^{+3.4+15.4}_{-2.7-2.5}$ & (0.3, 0.7) & 3 
& \settowidth{\LL}{300 arcmin$^2$}
\multirow{4}{\LL}{300 arcmin$^2$}&
\settowidth{\LL}{$Z \sim 25.9$}
\multirow{4}{\LL}{$Z \sim 25.9$}& 
\settowidth{\LL}{76 \%}
\multirow{4}{\LL}{76 \%} & \citet{dahlen2004}  
\\ 
$[0.6, 1.0]$ & \multicolumn{1}{c}{-} & $15.7^{+4.4+7.5}_{-2.5-5.3}$ & (0.3, 0.7) & 14 
& & & & \citet{dahlen2004}  
\\
$[1.0, 1.4]$ & \multicolumn{1}{c}{-} & $11.5^{+4.7+3.2}_{-2.6-4.4}$ & (0.3, 0.7) & 6 
& & & & \citet{dahlen2004}  
\\
$[1.4, 1.8]$ & \multicolumn{1}{c}{-} & $4.4^{+3.2+1.4}_{-2.5-1.1}$ & (0.3, 0.7) & 2 
& & & & \citet{dahlen2004}  
\\

$[0.2, 0.3]$ & & $1.7\pm 1.7$ & (0.3, 0.7) & 1 
& \settowidth{\LL}{2.3 deg$^2$}
\multirow{6}{\LL}{2.3 deg$^2$}&
\settowidth{\LL}{$m \sim 24.2$}
\multirow{6}{\LL}{$m \sim 24.2$}& 
\settowidth{\LL}{24 \%}
\multirow{6}{\LL}{24 \%} & \citet{barris2006}$^d$  
\\ 
$[0.3, 0.4]$ & \multicolumn{1}{c}{-} & $5.3\pm 2.4$ & (0.3, 0.7) & 5 
& & & & \citet{barris2006}$^d$  
\\
$[0.4, 0.5]$ & \multicolumn{1}{c}{-} & $7.3\pm 2.4$ & (0.3, 0.7) & 9 
& & & & \citet{barris2006}$^d$  
\\
$[0.5, 0.6]$ & \multicolumn{1}{c}{-} & $20.4\pm 3.8$ & (0.3, 0.7) & 29 
& & & & \citet{barris2006}$^d$  
\\
$[0.6, 0.7]$ & \multicolumn{1}{c}{-} & $14.9\pm 3.1$ & (0.3, 0.7) & 23 
& & & & \citet{barris2006}$^d$  
\\
$[0.7, 0.8]$ & \multicolumn{1}{c}{-} & $17.8\pm 3.4$ & (0.3, 0.7) & 28 
& & & & \citet{barris2006}$^d$  
\\
0.47         & $0.154^{+0.039+0.048}_{-0.031-0.033}$ &
$4.2^{+0.6+1.3}_{-0.6-0.9}$    & (0.3, 0.7) & 73 
& 4 deg$^2$& $i' \sim 23.3$ & 100 \% & \citet{neill2006}$^f$  \\
$[0.0, 0.5]$ & \multicolumn{1}{c}{-} & $0.0^{+2.4}_{-0.0}$ & (0.3, 0.7) & 0.0 
& \settowidth{\LL}{900 arcmin$^2$}
\multirow{4}{\LL}{900 arcmin$^2$}&
\settowidth{\LL}{$z' = 26.3$}
\multirow{4}{\LL}{$z' = 26.3$}& 
\settowidth{\LL}{0 \%}
\multirow{4}{\LL}{0 \%} & \citet{poznanski2007}$^g$
\\ 
$[0.5, 1.0]$ & \multicolumn{1}{c}{-} & $4.3^{+3.6}_{-3.2}$ & (0.3, 0.7) & 5.5 
& & & & \citet{poznanski2007}$^g$
\\
$[1.0, 1.5]$ & \multicolumn{1}{c}{-} & $10.5^{+4.5}_{-5.6}$ & (0.3, 0.7) & 10 
& & & & \citet{poznanski2007}$^g$
\\
$[1.5, 2.0]$ & \multicolumn{1}{c}{-} & $8.1^{+7.9}_{-6.0}$ & (0.3, 0.7) & 3 
& & & & \citet{poznanski2007}$^g$
\\



$[0.2, 0.6]$ & \multicolumn{1}{c}{-} & $5.3^{+3.9}_{-1.7}$ & (0.3, 0.7) & 5.44 
& \settowidth{\LL}{300 arcmin$^2$}
\multirow{4}{\LL}{300 arcmin$^2$}&
\settowidth{\LL}{$Z \sim 25.9$}
\multirow{4}{\LL}{$Z \sim 25.9$}& 
\settowidth{\LL}{76 \%}
\multirow{4}{\LL}{76 \%} & \citet{kuznetsova2008}$^{g,h}$
\\ 
$[0.6, 1.0]$ & \multicolumn{1}{c}{-} & $9.3^{+2.5}_{-2.5}$ & (0.3, 0.7) & 18.33 
& & & & \citet{kuznetsova2008}$^{g,h}$
\\
$[1.0, 1.4]$ & \multicolumn{1}{c}{-} & $7.5^{+3.5}_{-3.0}$ & (0.3, 0.7) & 8.87
& & & & \citet{kuznetsova2008}$^{g,h}$
\\
$[1.4, 1.7]$ & \multicolumn{1}{c}{-} & $1.2^{+5.8}_{-1.2}$ & (0.3, 0.7) & 0.35 
& & & & \citet{kuznetsova2008}$^{g,h}$\\
0.30 & \multicolumn{1}{c}{-} & $3.4^{+1.6+2.1}_{-1.5-2.2}$ & (0.3, 0.7) & 86 & 43000 gal & $R \sim
23$ &  30 \% & \citet{botticella2008}
\end{tabular}
\caption{Published restframe type~Ia supernova explosion rate
measurements. These rates are given together with the mean redshift of
the observed SNe and the cosmological model assumed (especially for
distant values). The fifth to eighth columns give respectively: the
number of supernovae from which the rate is computed; the effective
surface of the survey, or the number of galaxies surveyed; the
limiting magnitude of the survey and the fraction of spectroscopically
confirmed supernovae. \textsc{Notes}: $a$)~Quoted error bar are first
statistics, and second systematics; where only one is present, the
systematic is not available; $b$)~the value per comoving volume is
derived from the value in SNu times the 2dF luminosity density (the
error on the luminosity density is added quadratically to the
systematic uncertainty if distinct); $c$)~For a (0.3,~0.7) cosmology
the rate decreases by less than 5~\%~\protect\citep{hardin2000};
$d$)~\protect\cite{barris2006} do not provide systematic errors on
their measurements; $f$)~\protect\cite{neill2006} used the
\citet{ilbert2005} luminosity function to translate their rate in SNu;
$g$)~statistical and systematic errors are added in quadrature;
$h$)~\protect\cite{kuznetsova2008} include the SN sample from
\citet{dahlen2004} to compute their rate.}
\label{tab:otheresultsIa}

\end{table*}

%
\begin{table*}[!ht]
\centering
\begin{tabular}{lllcccccl}
\hline
      & \multicolumn{2}{c}{$R_{\mathrm{SN II/Ibc}}$} 
&  & SNe &  survey& limiting & spectro & \\
\cline{2-3}
\raisebox{2.5ex}[0cm][0cm]{$\langle z\rangle$}  & ($h_{70}^2$ SNu) & 
($10^{-5}\ h_{70}^3\
\mathrm{Mpc}^{-3}\ \mathrm{yr}^{-1}$) & 
\raisebox{2.5ex}[0cm][0cm]{($\Omega_{M_{\circ}}, 
\Omega_{\Lambda_{\circ}}$)} & 
nb &
surface &
mag &
confirmed &
\multicolumn{1}{c}{\raisebox{2.5ex}[0cm][0cm]{author}} \\ 
\hline
$\sim$ 0    & $0.41\pm0.17$ & 4.8$\pm$1.9 & & 67 &
$10^4$ gal& $R \sim 20$ & 100 \% 
& \citet{cappellaro1999}$^a$ \\
0.26 & $1.26^{+0.48}_{-0.39}$ & $19.2^{+7.0}_{-6.1}$ & (0.3, 0.7) &
28.2 & 11300 gal & $V \sim 24.5$ & 20 \% 
& \citet{cappellaro2005} \\
$[0.1, 0.5[$ & & $25.1^{+8.8+7.5}_{-7.5-18.6}$ & (0.3, 0.7) & 6 
& \settowidth{\LL}{300 arcmin$^2$}
\multirow{2}{\LL}{300 arcmin$^2$}&
\settowidth{\LL}{$Z \sim 25.9$}
\multirow{2}{\LL}{$Z \sim 25.9$}& 
\settowidth{\LL}{63 \%}
\multirow{2}{\LL}{63 \%}
& \citet{dahlen2004}$^b$ \\
$[0.5, 0.9[$ & & $39.6^{+10.3+19.2}_{-10.6-26.0}$ & (0.3, 0.7) & 10
& & & 
&    \citet{dahlen2004}$^b$ \\
0.21 & & $11.5^{+4.3+4.2}_{-3.3-3.6}$ & (0.3, 0.7) & 86 & 43000 gal & $R \sim
23$ &  30 \% & \citet{botticella2008}$^{b,c}$
\end{tabular}
\caption{Published restframe CCSN explosion rate measurements. These
rates are given together with the mean redshift or redshift range of
the observed SNe and the cosmological model assumed. The fifth to
eighth columns give respectively: the number of supernovae from which
the rate is computed; the effective surface of the survey, or the
number of galaxies surveyed; the limiting magnitude of the survey and
the fraction of spectroscopically confirmed
supernovae. \textsc{Notes}: $a$)~The rate is translated into
volumetric units by using a luminosity density of galaxies as in
\citet{cappellaro2005} $b$)~We give here the value corrected for
extinction by the authors $c$)~\protect\cite{botticella2008} include
the SN sample from \citet{cappellaro2005} to compute their rate.}
\label{tab:otheresultsCC}
\end{table*}
%

\end{document}